\documentclass[pre,twocolumn,amsmath,amssymb,floatfix,superscriptaddress]{revtex4}

\usepackage{graphicx}
\usepackage{latexsym}
\usepackage{amsmath}
\usepackage{amssymb}
\usepackage{amsfonts}
\usepackage{bm}

\newcommand{\la}{\left<}
\newcommand{\ra}{\right>}
\newcommand{\fvec}{\ensuremath{\underline{f}}}

\newcommand{\nvecl}{\underline{n}_l}
\newcommand{\pvec}{\ensuremath{\underline{p}}}
\newcommand{\qvec}{\ensuremath{\underline{q}}}
\newcommand{\rvec}{\ensuremath{\underline{r}}}
\newcommand{\rvecl}{\ensuremath{\underline{r}_l}}
\newcommand{\uvec}{\ensuremath{\underline{u}}}
\newcommand{\vvec}{\ensuremath{\underline{v}}}
\newcommand{\Mmat}{\underline{\underline{M}}}
\newcommand{\ddiff}{\ensuremath{\text{d}}}

\newcommand{\nlx}{n_{l,x}}
\newcommand{\nly}{n_{l,y}}
\newcommand{\ul}{u_l}
\newcommand{\rl}{r_l}
\newcommand{\rx}{r_x}
\newcommand{\ry}{r_y}

\newcommand{\vix}{v_{i,x}}
\newcommand{\viy}{v_{i,y}}
\newcommand{\px}{p_x}
\newcommand{\py}{p_y}

\newcommand{\kB}{\mbox{$k_{\rm B}$}}
\newcommand{\kBT}{\mbox{$k_{\rm B}T$}}
\newcommand{\NVgT}{\ensuremath{\text{NV}\gamma\text{T}}}
\newcommand{\NVtT}{\ensuremath{\text{NV}\tau\text{T}}}

\newcommand{\Pid}{\ensuremath{P_\mathrm{id}}}
\newcommand{\Pex}{\ensuremath{P_\mathrm{ex}}}

\newcommand{\muA}{\ensuremath{\mu_\mathrm{A}}}
\newcommand{\muAhat}{\ensuremath{\hat{\mu}_\mathrm{A}}}
\newcommand{\muAid}{\ensuremath{\mu_\mathrm{A,id}}}
\newcommand{\muAidhat}{\ensuremath{\hat{\mu}_\mathrm{A,id}}}
\newcommand{\muAex}{\ensuremath{\mu_\mathrm{A,ex}}}
\newcommand{\muAexhat}{\ensuremath{\hat{\mu}_\mathrm{A,ex}}}
\newcommand{\muB}{\ensuremath{\mu_\mathrm{B}}}
\newcommand{\muF}{\ensuremath{\sigma_\mathrm{F}}}
\newcommand{\muFid}{\ensuremath{\sigma_\mathrm{F,id}}}
\newcommand{\muFex}{\ensuremath{\sigma_\mathrm{F,ex}}}
\newcommand{\muFtau}{\ensuremath{\left.\sigma_\mathrm{F}}\right|_{\tau}}
\newcommand{\muFgam}{\ensuremath{\left.\sigma_\mathrm{F}}\right|_{\gamma}}
\newcommand{\Cttau}{\ensuremath{\left.C(t)\right|_{\tau}}}
\newcommand{\Ctgam}{\ensuremath{\left.C(t)\right|_{\gamma}}}

\newcommand{\Keq}{K_\mathrm{eq}}
\newcommand{\Meq}{M_\mathrm{eq}}
\newcommand{\Geq}{G_\mathrm{eq}}
\newcommand{\GM}{G_\mathrm{M}}

\newcommand{\Gstor}{G^{\prime}(\omega)}
\newcommand{\Gloss}{G^{\prime\prime}(\omega)}
\newcommand{\Cstor}{C^{\prime}(\omega)}
\newcommand{\Closs}{C^{\prime\prime}(\omega)}
\newcommand{\Mstor}{M^{\prime}(\omega)}
\newcommand{\Mloss}{M^{\prime\prime}(\omega)}

\newcommand{\tauhat}{\hat{\tau}}
\newcommand{\sighat}{\hat{\sigma}}
\newcommand{\tauid}{\tau_\mathrm{id}}
\newcommand{\tauidhat}{\hat{\tau}_\mathrm{id}}
\newcommand{\tauex}{\tau_\mathrm{ex}}
\newcommand{\tauexhat}{\hat{\tau}_\mathrm{ex}}
\newcommand{\Ihat}{\hat{I}}

\newcommand{\Ahat}{\ensuremath{\hat{A}}}
\newcommand{\Bhat}{\ensuremath{\hat{B}}}
\newcommand{\dAhat}{\ensuremath{\delta\hat{A}}}
\newcommand{\dBhat}{\ensuremath{\delta\hat{B}}}
\newcommand{\Hhat}{\hat{\cal H}}
\newcommand{\Hidhat}{\hat{\cal H}_\mathrm{id}}
\newcommand{\Hexhat}{\hat{\cal H}_\mathrm{ex}}
\newcommand{\gamAmp}{\gamma_0}
\newcommand{\dtMD}{\delta t_\mathrm{MD}}

\newcommand{\tauref}{\tau_\text{ref}}
\newcommand{\Tperiod}{T_\mathrm{\omega}}

\newcommand{\thetae}{\theta_{1/e}}
\newcommand{\thetaone}{\theta_{1\%}}
\newcommand{\thetafif}{\theta_{5\%}}
\newcommand{\thetaG}{\tilde{\theta}}
\newcommand{\ctrans}{\ensuremath{c_{\bot}}}
\newcommand{\utrans}{\ensuremath{\underline{u}_{\bot}}}

\begin{document}

\title{Fluctuation-dissipation relation between shear stress relaxation\\ 
modulus and shear stress autocorrelation function revisited}

\author{J.P.~Wittmer}
\affiliation{Institut Charles Sadron, Universit\'e de Strasbourg \& CNRS, 23 rue du Loess, 67034 Strasbourg Cedex, France}
\author{H.~Xu}
\email{hongxu@univ-lorraine.fr}
\affiliation{LCP-A2MC, Institut Jean Barriol, Universit\'e de Lorraine \& CNRS, 1 bd Arago, 57078 Metz Cedex 03, France}
\author{O. Benzerara}
\affiliation{Institut Charles Sadron, Universit\'e de Strasbourg \& CNRS, 23 rue du Loess, 67034 Strasbourg Cedex, France}
\author{J. Baschnagel}
\affiliation{Institut Charles Sadron, Universit\'e de Strasbourg \& CNRS, 23 rue du Loess, 67034 Strasbourg Cedex, France}

\begin{abstract}
The shear stress relaxation modulus $G(t)$ may be determined from the shear stress $\tauhat(t)$ after 
switching on a tiny step strain $\gamma$ or by inverse Fourier transformation of the storage modulus $\Gstor$
or the loss modulus $\Gloss$ obtained in a standard oscillatory shear experiment at angular frequency $\omega$.
It is widely assumed that $G(t)$ is equivalent in general to the equilibrium stress 
autocorrelation function $C(t) = \beta V \langle \delta \tauhat(t) \delta \tauhat(0)\rangle$ 
which may be readily computed in computer simulations
($\beta$ being the inverse temperature and $V$ the volume).
Focusing on isotropic solids formed by permanent spring networks
we show theoretically by means of the fluctuation-dissipation theorem
and computationally by molecular dynamics simulation that in general
$G(t) = \Geq + C(t)$ for $t > 0$ with $\Geq$ being the static equilibrium shear modulus.
A similar relation holds for $\Gstor$.
$G(t)$ and $C(t)$ must thus become different for a solid body 
and it is impossible to obtain $\Geq$ directly from $C(t)$. 
\end{abstract}
\date{\today}
\maketitle

\section{Introduction}
\label{sec_intro}

\begin{figure}[t]
\centerline{\resizebox{0.9\columnwidth}{!}{\includegraphics*{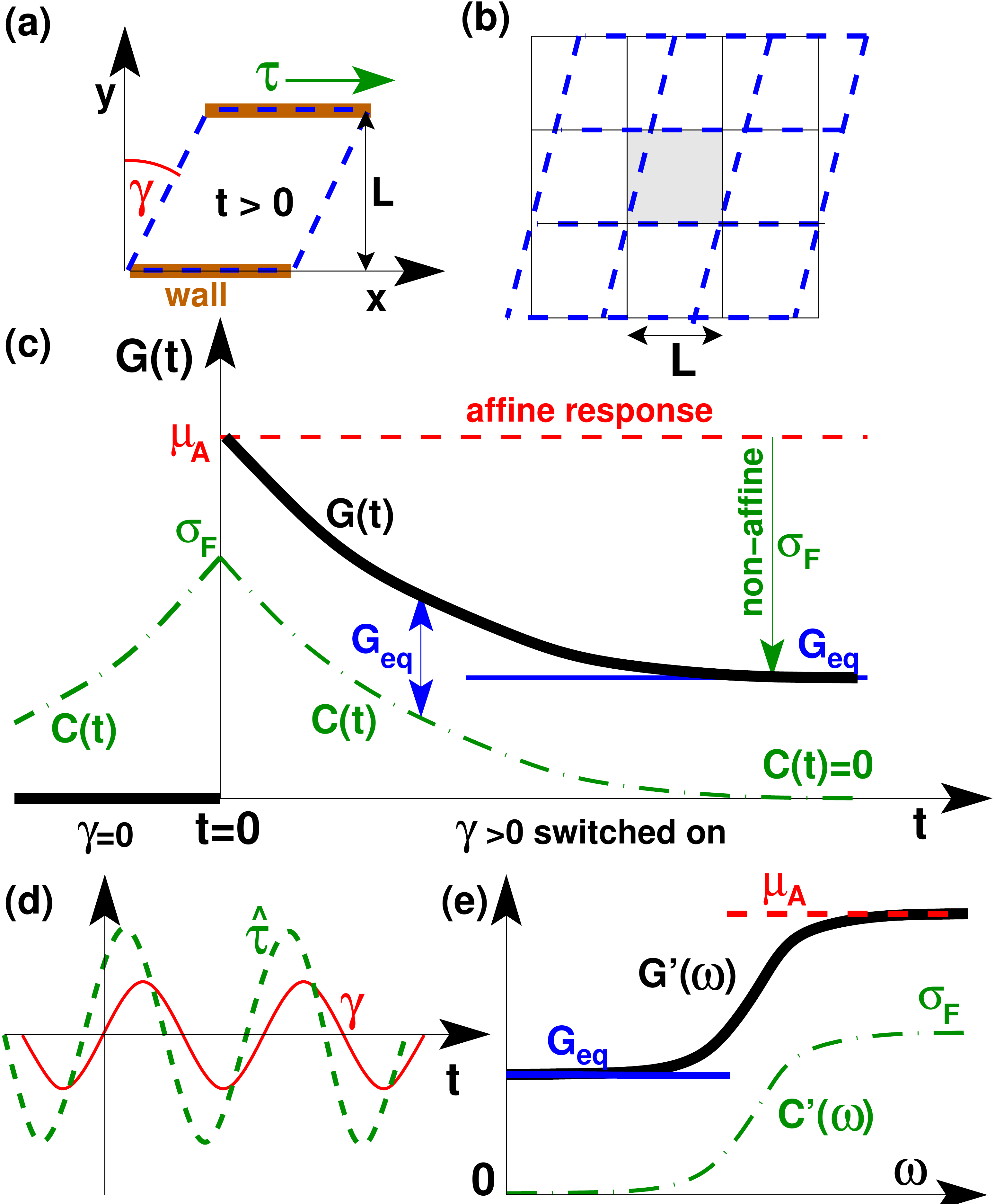}}}
\caption{Sketch of notations and addressed problem:
{\bf (a)} experimental setup,
{\bf (b)} plain shear with periodic boundary conditions,
{\bf (c)} sketch of key equation, Eq.~(\ref{eq_key}),
with $\muF = \beta V \langle \delta \tauhat^2 \rangle$ characterizing the stress fluctuations
at $\gamma=0$,
{\bf (d)} imposed sinusoidal shear strain $\gamma(t)$ (solid line)
and measured shear stress $\tauhat(t)$ (dashed line),
{\bf (e)} comparison of $\Cstor$ (dash-dotted line) and storage modulus $\Gstor$, Eq.~(\ref{eq_key_Gstor}).
\label{fig_sketch}
}
\end{figure}

\paragraph*{Background.}
A central rheological property characterizing the linear response of liquids and solid elastic bodies
is the shear relaxation modulus $G(t)$ \cite{RubinsteinBook,WittenPincusBook,DoiEdwardsBook}.
Assuming for simplicity an isotropic system, the shear relaxation modulus $G(t) = \delta \tau(t)/\gamma$
may be obtained from the stress increment $\delta \tau(t) = \la \tauhat(t) - \tauhat(0^{-}) \ra$ for $t > 0$
after a small step strain with $|\gamma| \ll 1$ has been imposed at time $t=0$ as sketched in panel (a) of
Fig.~\ref{fig_sketch}. 
The instantaneous shear stress $\tauhat(t)$ at time $t$ may be determined experimentally by probing the 
forces acting on the walls of the shear cell and in a numerical study,
as shown in panel (b) for a sheared periodic simulation box,
from the imposed model Hamiltonian and the particle positions and momenta 
\cite{AllenTildesleyBook,FrenkelSmitBook,ThijssenBook,LandauBinderBook}.
It is well known that the components of the Fourier transformed relaxation modulus $G(t)$,
the storage modulus $\Gstor$ and the loss modulus $\Gloss$, are directly measurable in 
an oscillatory shear strain experiment \cite{RubinsteinBook} as shown in panel (d) and panel (e).
Using either $G(t)$ or $\Gstor$ one obtains in the static limit 
(bold horizontal solid lines) the shear modulus \cite{RubinsteinBook,LandauElasticity}
\begin{equation}
\Geq = \lim_{t\to \infty} G(t) = \lim_{\omega \to 0} \Gstor.
\label{eq_Geqdef}
\end{equation}
The shear modulus is an important order parameter characterizing the
transition from the liquid/sol ($\Geq=0$) to the solid/gel state ($\Geq > 0)$ where
the particle permutation symmetry of the liquid state is lost on the time window probed
\cite{WittenPincusBook,Alexander98}.
Examples of current interest for the determination of $\Geq$ include 
crystalline solids \cite{Biroli10},
glass forming liquids and amorphous solids
\cite{GoetzeBook,Barrat88,WTBL02,TWLB02,Berthier05,Berthier07,Mezard10,Szamel11,Yoshino12,Klix12,WXP13,ZT13,WXP13c,Barrat13,Szamel15,WXB15},
colloidal gels \cite{Kob08}, 
polymeric networks \cite{RubinsteinBook,DKG91,DKG94,Zippelius06},
hyperbranched polymer chains with sticky end-groups \cite{Friedrich10} or
bridged equilibrium networks of telechelic polymers \cite{Porte03}.

\paragraph*{Key issue.}
Surprisingly, it is widely assumed \cite{GoetzeBook,DKG91,DKG94,Klix12,Szamel15} 
that in the linear response limit ($\gamma \to 0$) the stress relaxation modulus $G(t)$ 
must become equivalent to the {\em stress autocorrelation function} \cite{AllenTildesleyBook} 
$C(t) \equiv \beta V \la \delta \tauhat(t) \delta \tauhat(0) \ra$
computed in the \NVgT-ensemble at imposed particle number $N$, volume $V$, shear strain $\gamma$ and 
temperature $T$ ($\beta = 1/\kBT$ denoting the inverse temperature) \cite{foot_gamnot}.
Since $G(t)=C(t)$ is assumed to hold, $\Geq$ is supposed to be measurable from some transient 
``finite frozen-in amplitude" of $C(t)$ \cite{Klix12}.
We call this belief ``surprising" since, obviously, $C(t) \to 0$ in the thermodynamic limit
\cite{HansenBook}
for large times $t \gg \theta$ with $\theta$ characterizing the typical relaxation time
of a shear stress fluctuation (properly defined in Sec.~\ref{simu_key}).
At variance to this belief we shall show by inspection of 
the fluctuation-dissipation theorem (FDT) \cite{HansenBook} that more generally 
\begin{equation}
G(t) - \Geq = C(t) \ \mbox{ for } t \ge 0
\label{eq_key}
\end{equation}
and $G(t)=0$ for $t <0$ as sketched in panel (c).
Two immediate consequences of Eq.~(\ref{eq_key}) are 
\begin{enumerate}
\item
that $G(t)$ only becomes equivalent to $C(t)$ for $t > 0$ in the liquid limit where (trivially) $\Geq=0$
and 
\item
that the shear modulus $\Geq$ is only probed by $G(t)$ on time scales $t \gg \theta$ where $C(t)$ 
must vanish.  
\end{enumerate}
In principle, it is thus {\em impossible} to obtain the static shear modulus $\Geq$ 
of an elastic body only from $C(t)$.
We shall show that a similar relation 
\begin{eqnarray}
\Gstor - \Geq & = & \Cstor \equiv \int_0^{\infty} C(t) \sin(\omega t) \ddiff (\omega t) \label{eq_key_Gstor} \\
\Gloss & = & \Closs \equiv \int_0^{\infty} C(t) \cos(\omega t) \ddiff (\omega t) \label{eq_key_Gloss} 
\end{eqnarray}
holds in the angular frequency domain as sketched in panel (e) of Fig.~\ref{fig_sketch} for $\Gstor$.
We refer below to Eqs.~(\ref{eq_key}-\ref{eq_key_Gloss}) as the ``key relations". 

\paragraph*{Outline.}
The presented work is closely related to the recent paper \cite{WXB15} where we focus on the 
difference of static fluctuations and autocorrelation functions in conjugated ensembles
and where we also discuss briefly transient self-assembled networks.
The present paper provides complementary informations focusing on permanent elastic networks
in the $\NVgT$-ensemble and on the response to an oscillatory shear strain 
$\gamma(t) = \gamAmp \sin(\omega t)$.
We begin by reminding in Sec.~\ref{theo_static} the ``affine" and ``stress fluctuation" contributions 
$\muA$ and $\muF$ to the equilibrium shear modulus $\Geq = \muA - \muF$ and demonstrate 
then the key relations theoretically.
In Sec.~\ref{sec_model} we define our two-dimensional spring model and give 
some algorithmic details. The construction of the network and some properties of its 
athermal ground state are presented in Sec.~\ref{sec_refer}. 
Our computational results for one finite temperature are given in Sec.~\ref{sec_simu}.
Some static properties are summarized in Sec.~\ref{simu_static}.
We illustrate numerically in Sec.~\ref{simu_key} that Eq.~(\ref{eq_key}) holds.
We focus in Sec.~\ref{simu_oscil} on the storage and loss moduli $\Gstor$ and $\Gloss$ 
computed directly by applying a sinusoidal shear strain and compare this result to the
Fourier transformations of $G(t)$ and $C(t)$.
We summarize our work in Sec.~\ref{sec_conc} where we briefly comment on 
the generalization of the key relations to linear response functions of other intensive variables.

\section{Theoretical considerations}
\label{sec_theo}

\subsection{Reminder of some static properties}
\label{theo_static}

\paragraph*{Affine canonical displacements.}
Let us consider first an infinitessimal pure shear strain $\gamma$ of the periodic simulation 
box in the $(x,y)$-plane as sketched in panel (b) of Fig.~\ref{fig_sketch}.
We assume that not only the box shape is changed but that the particle positions $\rvec$
(using the principal box convention) follow the macroscopic constraint in an
{\em affine} manner according to 
\begin{equation}
\rx \to \rx + \gamma \ry \ \mbox{ for } |\gamma| \ll 1.
\label{eq_r_cantrans}
\end{equation}
Albeit not strictly necessary for the demonstration of the key relations we focus on, 
we shall assume, moreover, that this shear transformation is also {\em canonical} 
\cite{Goldstein,FrenkelSmitBook}. This implies that the momenta must transform as 
\begin{equation}
\px \to \px - \gamma \py \ \mbox{ for } |\gamma| \ll 1.
\label{eq_p_cantrans}
\end{equation}
All other coordinates of the positions and momenta remain unchanged in the presented case 
\cite{foot_affinecanonic}. We emphasize the negative sign in Eq.~(\ref{eq_p_cantrans}) 
which assures that Liouville's theorem is obeyed \cite{Goldstein}. 

\paragraph*{Instantaneous shear stress and affine shear elasticity.}
Let $\Hhat(\gamma)$ denote the system Hamiltonian of a given state $s$ written 
as a function of the shear strain $\gamma$. 
The instantaneous shear stress $\tauhat$ and the instantaneous affine shear elasticity $\muAhat$
may be defined as the expansion coefficients associated to the energy change
\begin{equation}
\delta \Hhat(\gamma)/V = \tauhat \gamma + \muAhat \gamma^2/2 \mbox{ for } |\gamma| \ll 1
\label{eq_Haffine}
\end{equation}
with $\gamma=0$ being the reference, i.e. \cite{foot_prime}
\begin{eqnarray}
\tauhat & \equiv & \Hhat^{\prime}(\gamma)/V|_{\gamma=0} \label{eq_tauhatdef} \mbox{ and } \\
\muAhat & \equiv & \Hhat^{\prime\prime}(\gamma)/V|_{\gamma=0} \label{eq_muAhatdef}.
\end{eqnarray}
With $\Hidhat(\gamma)$ and $\Hexhat(\gamma)$ being the standard
ideal kinetic and the (conservative) excess interaction contributions
to the total Hamiltonian $\Hhat(\gamma) = \Hidhat(\gamma) + \Hexhat(\gamma)$,
this implies similar relations for the corresponding contributions
$\tauidhat$ and $\tauexhat$ to  $\tauhat =\tauidhat + \tauexhat$ and 
for the contributions $\muAidhat$ and $\muAexhat$ to $\muAhat = \muAidhat + \muAexhat$. 
The ideal contributions $\tauidhat$ and $\muAidhat$ are then due to the change
of $\Hidhat(\gamma)$ imposed by the momentum transform Eq.~(\ref{eq_p_cantrans}),
the excess contributions $\tauexhat$ and $\muAexhat$ due the change of
the $\Hexhat(\gamma)$ imposed by the strained particle positions, Eq.~(\ref{eq_r_cantrans}).

\paragraph*{Ideal contributions.}
Using Eq.~(\ref{eq_p_cantrans}) and $\pvec_i = m_i \vvec_i$ with $m_i$ being the mass of 
particle $i$ and $\vvec_i$ its velocity, the kinetic energy of the strained system
becomes $\Hidhat(\gamma) = \sum_i m_i [ (\vix- \gamma \viy)^2 + \viy^2 \ldots]/2$.
This implies that
\begin{eqnarray}
\tauidhat & = & - \frac{1}{V} \sum_{i=1}^N m_i \vix \viy \label{eq_tauidhat} \mbox{ and } \\
\muAidhat & = & \frac{1}{V} \sum_{i=1}^N m_i \viy^2 \label{eq_muAidhat} 
\end{eqnarray}
for the ideal contributions to the shear stress and the affine shear elastiticy.
Note that the minus sign for the shear stress is due to the minus sign in Eq.~(\ref{eq_p_cantrans})
required for a canonical transformation.

\paragraph*{Excess contributions.}
We focus below on pairwise additive excess energies
$\Hexhat = \sum_l u(\rl)$ with $u(r)$ being a pair potential and where the 
running index $l$ labels the interaction between two particles $i < j$.
(The potential $u(r)$ may in addition explicitly depend on the interaction $l$.)
Due to Eq.~(\ref{eq_r_cantrans}) a squared particle distance $r^2=x^2+y^2+\ldots$ becomes 
$r(\gamma)^2 = (x + \gamma y)^2 + y^2 + \ldots$ 
Straightforward application of the chain rule \cite{WXP13} shows for
the excess contributions to $\tauhat$ and $\muA$ that
\begin{eqnarray}
\tauexhat & = & \frac{1}{V} \sum_l \rl u^{\prime}(\rl) \ \nlx \nly   \label{eq_tauexhat} \ \mbox{ and } \\
\muAexhat & = & \frac{1}{V} \sum_l  \left( \rl^2 u^{\prime\prime}(\rl)
- \rl u^{\prime}(\rl) \right) \nlx^2 \nly^2 \nonumber \\
& + & \frac{1}{V} \sum_l \rl u^{\prime}(\rl) \ \nly^2  \label{eq_muAexhat}
\end{eqnarray}
with $\nvecl = \rvecl/\rl$ being the normalized distance vector $\rvec= \rvec_j - \rvec_i$
between the particles $i$ and $j$. 
Interestingly, Eq.~(\ref{eq_tauexhat}) is strictly identical to the 
corresponding off-diagonal term of the Kirkwood stress tensor  \cite{AllenTildesleyBook}.

\paragraph*{Thermodynamic averages.}
Let us assume an isotropic elastic body at imposed particle number $N$, constant volume $V$,
shear strain $\gamma$ and mean temperature $T$ ($\NVgT$-ensemble).
Note that the (intensive) shear strain $\gamma$ corresponds thermodynamically 
to an {\em extensive} variable $X = \gamma V$. We write $F(\gamma)$ for the free energy and
\begin{equation}
Z(\gamma) = \exp( -\beta F(\gamma)) = \sum_s \exp(- \beta \Hhat(\gamma))
\label{eq_Zgamma}
\end{equation}
for the corresponding partition function with $\sum_s$ being the sum over all accessible system states. 
Thermal averages are given by 
$\langle \ldots \rangle = \sum_s \ldots \exp(- \beta \Hhat(\gamma))/Z(\gamma)$.
Interestingly, the definition of the instantaneous shear stress $\tauhat$ given above,
Eq.~(\ref{eq_Haffine}), is consistent with the thermodynamic mean shear stress 
$\tau \equiv \langle \tauhat \rangle$,
while the average affine shear elasticity $\muA \equiv \langle \muAhat \rangle$ 
only corresponds to an {\em upper bound} to the shear modulus $\Geq$.
To see this let us first note for convenience that \cite{foot_prime}
\begin{eqnarray}
\frac{\partial \log(Z(\gamma))}{\partial \gamma} & = & \frac{Z^{\prime}(\gamma)}{Z(\gamma)} \label{eq_Zone} \\
\frac{\partial^2\log(Z(\gamma))}{\partial \gamma^2} & = & \frac{Z^{\prime\prime}(\gamma)}{Z(\gamma)} -
\left(\frac{Z^{\prime}(\gamma)}{Z(\gamma)}\right)^2 \label{eq_Ztwo} 
\end{eqnarray}
and
\begin{eqnarray}
Z^{\prime}(\gamma) & = & - \sum_s \beta \Hhat^{\prime}(\gamma) e^{-\beta \Hhat(\gamma)} \label{eq_Zthree} \\
Z^{\prime\prime}(\gamma) & = & \sum_s \left( (\beta \Hhat^{\prime}(\gamma))^2
- \beta \Hhat^{\prime\prime}(\gamma) \right) e^{-\beta \Hhat(\gamma)}. \label{eq_Zfour} 
\end{eqnarray}
It then follows using Eq.~(\ref{eq_Zone}) and Eq.~(\ref{eq_Zthree}) 
that the mean shear stress is indeed
\cite{WXP13}
\begin{eqnarray}
\tau &= & \left.\frac{\partial F(X)}{\partial X}\right|_{X=0} \label{eq_tauTD} \\
& = & \sum_s \Hhat^{\prime}(\gamma)/V|_{\gamma=0} \frac{\exp(-\beta \Hhat)}{Z} = \la \tauhat \ra.
\label{eq_avtau}
\end{eqnarray}
Using Eq.~(\ref{eq_Ztwo}) and Eq.~(\ref{eq_Zfour}) one verifies also that \cite{Callen}
\begin{eqnarray}
\Geq & = &  V \ \left.\frac{\partial^2 F(X)}{\partial X^2} \right|_{X=0} \label{eq_GeqTD} \\
& = & \muA - \muF \mbox{ with } \muF \equiv \beta V \la \delta \tauhat^2 \ra \label{eq_GeqNVgT}
\end{eqnarray}
characterizing the shear-stress fluctuations at $\gamma=0$.
Since $\muF \ge 0$, $\muA$ is only an upper bound to $\Geq$.
We emphasize that Eq.~(\ref{eq_GeqNVgT}) is a special case of the general stress-fluctuation 
relations for elastic moduli \cite{Hoover69,Barrat88,Lutsko89,Barrat13}. It provides a 
computational convenient method to obtain $\Geq$ for systems in the $\NVgT$-ensemble
used in many recent numerical studies \cite{Barrat88,WTBL02,TWLB02,WXP13,Szamel15,WXB15}.
We have demonstrated Eq.~(\ref{eq_GeqNVgT}) {\em without} 
introducing a local displacement field as in Ref.~\cite{Hoover69}.
We note {\em en passant} that the averages $\tau$ and $\muA$ are 
``simple averages", i.e. no fluctuations, and are thus identical for any 
ensemble given that the same state point is sampled 
\cite{AllenTildesleyBook,WXP13}.

\paragraph*{Lebowitz-Percus-Verlet transform.}
Equation~(\ref{eq_GeqNVgT}) can alternatively be obtained from the 
general transformation relation for a fluctuation $\langle \dAhat \dBhat \rangle$ of 
two observables ${\cal A}$ and ${\cal B}$ due to Lebowitz, Percus and Verlet 
\cite{Lebowitz67}
\begin{equation}
\left. \la \dAhat \dBhat \ra\right|_{I} =
\left. \la \dAhat \dBhat \ra\right|_{X}  
+ \frac{\partial (\beta I)}{\partial X} \frac{\partial \langle \Ahat \rangle}{\partial (\beta I)} 
\frac{\partial \langle \Bhat \rangle}{\partial (\beta I)}
\label{eq_dAdB}
\end{equation}
with $X = V\gamma$ denoting again the extensive variable and $I = \tau$ the conjugated intensive variable
\cite{foot_LPVsimplified}.
This gives 
\begin{equation}
\muFtau = \muFgam +  \Geq 
\label{eq_muFtaumuFgam}
\end{equation}
i.e. the thermodynamic shear modulus $\Geq$ compares the shear stress fluctuations
in the conjugated ensembles at constant mean shear stress $\tau$ and imposed shear strain $\gamma$.
The latter formula can be made more useful for computational studies by rewriting 
the shear stress fluctuations $\muFtau$ at constant shear stress $\tau$. 
Note that the normalized weight of a state $s$ in the \NVtT-ensemble is given by 
$p(\gamma) \sim \exp[-\beta (\Hhat(\gamma) -V\gamma \tau)]$. 
Using the instantaneous shear stress $\tauhat$ defined in Eq.~(\ref{eq_tauhatdef}) we thus have 
\begin{equation}
p^{\prime}(\gamma) = - \beta V [\tauhat(\gamma) - \tau)] p(\gamma).
\label{eq_pprime}
\end{equation}
Using integration by parts it is then readily seen \cite{WXP13}
that this leads to $\muFtau = \langle \Hhat^{\prime\prime}(\gamma)/V \rangle|_{\tau} 
= \langle \Hhat^{\prime\prime}(\gamma)/V \rangle|_{\gamma} = \muA$ 
in agreement with Eq.~(\ref{eq_muAhatdef}).
This confirms Eq.~(\ref{eq_GeqNVgT}) \cite{foot_gamnot}.

\paragraph*{Simplifications.}
Thermal averaging of Eq.~(\ref{eq_tauidhat}) and Eq.~(\ref{eq_muAidhat}) implies
$\tauid = \muAid = \Pid$ with $\Pid$ being the ideal normal pressure.
We note further that $\muF = \muFid + \muFex$ may again be rewritten as the 
sum of an ideal contribution $\muFid \equiv \beta V \langle \delta \tauidhat^2 \rangle = \Pid$ 
and an excess term $\muFex \equiv \beta V \langle \delta \tauexhat^2 \rangle$.
All ideal contributions to $\Geq$ thus cancel and one may rewrite Eq.~(\ref{eq_GeqNVgT}) as 
\begin{equation}
\Geq =  \muAex - \muFex.
\label{eq_Geqex}
\end{equation}
Since an ideal gas must have a vanishing shear modulus, this simplification is, of course, 
expected and could have been used from the start \cite{WXP13}.
Using Eq.~(\ref{eq_tauidhat}) and the fact that in an isotropic system all coordinates
are equivalent it is seen that the average shear elasticity reduces to \cite{WXP13}
\begin{eqnarray}
\muAex & = & \muB - \Pex   \mbox{ with } \nonumber \\
\muB & = & \frac{1}{V} \sum_l 
\la \left( \rl^2 u^{\prime\prime}(\rl) - \rl u^{\prime}(\rl) \right) \nlx^2 \nly^2 \ra
\label{eq_muA}
\end{eqnarray}
being the Born-Lam\'e term \cite{BornHuang,Lutsko89}
and $\Pex$ the excess contribution to the normal pressure $P=\Pid+\Pex$.

\subsection{Demonstration of Eq.~(\ref{eq_key})}
\label{theo_key}

\paragraph*{Static limits.}
As shown by the dash-dotted line in panel (c) of Fig.~\ref{fig_sketch}, 
by definition $C(t) \to \muF$ for $t \to 0$ and $C(t) \to 0$ for $t \to \infty$ \cite{HansenBook}. 
Equation~(\ref{eq_key}) thus implies that the relaxation modulus becomes 
\begin{equation}
G(t) \to \muF + (\muA-\muF) = \muA \mbox{ for } t \to 0^{+},
\label{eq_Gt2muA}
\end{equation}
which is consistent with an affine canonical shear strain imposed at $t=0$, Eq.~(\ref{eq_Haffine}),
and $G(t) \to \Geq$ for $t \to \infty$ as expected from Eq.~(\ref{eq_Geqdef}).

\paragraph*{Fluctuation Dissipation Theorem.}
We show next that Eq.~(\ref{eq_key}) must hold for all times.
Using Boltzmann's superposition principle for an arbitrary strain history $\gamma(t)$ \cite{RubinsteinBook} 
the shear stress becomes \cite{DoiEdwardsBook}
\begin{eqnarray}
\tau(t) & = & \int_{-\infty}^{t} \ddiff s \ G(t-s) \frac{d \gamma(s)}{d s} \label{eq_Boltzsup1} \\ 
        & = & \left. G(t-s) \gamma(s) \right|_{-\infty}^{t} - 
\int_{-\infty}^{t} \ddiff s \frac{d G(t-s)}{d s} \gamma(s) \nonumber 
\end{eqnarray} 
using integration by parts in the second step. 
Introducing the ``after-effect function" or ``dynamic response function" \cite{foot_prime}
$\chi(t) \equiv - G^{\prime}(t) = G^{\prime}(-t)$ 
this may be rewritten for a step strain imposed at $t=0$ as
\begin{equation}
G(t) = \Geq + \int_t^{\infty} \chi(s) \ \ddiff s \ 
\label{eq_Gtchit}
\end{equation}
being consistent with the expected $G(t) \to \Geq$ for $t \to \infty$.
Since according to the FDT as formulated by Eq.~(7.6.13) of Ref.~\cite{HansenBook},
the after-effect function is $\chi(t) = - C^{\prime}(t)$,
this yields the claimed relation Eq.~(\ref{eq_key}).

\paragraph*{Alternative demonstration.}
It is of some importance that Eq.~(\ref{eq_key}) may be also obtained from the general transformation 
relation Eq.~(\ref{eq_dAdB}) with $A = \tau(t)$ and $B = \tau(0)$.  It is assumed here that the 
shear-barostat imposing a mean shear stress $\tau = 0$ is sufficiently slow such that the system 
trajectory is not altered over the time scales used for the determination of the correlation functions 
\cite{Lebowitz67,AllenTildesleyBook,foot_Berthier}.
Generalizing the relation between the static fluctuations Eq.~(\ref{eq_muFtaumuFgam}) into the time domain, 
this yields immediately 
\begin{equation}
\Cttau = \Ctgam + \Geq.
\label{eq_CttauCtgam}
\end{equation}
As before for the static shear stress fluctuations $\muFtau$ one can show that
\begin{equation}
\Cttau = \left. \la \frac{\partial \tauhat(t;\gamma)}{\partial \gamma} \ra \right|_{\tau} = G(t) \mbox{ for } t 
> 0
\label{eq_GtCttau}
\end{equation}
where we have reexpressed in the first step $[\tauhat(t;\gamma)-\tau] [\tauhat(0;\gamma)-\tau] p(\gamma)$ 
using Eq.~(\ref{eq_pprime}). In the second step we have used that within linear response $G(t)$ 
does not depend on $\gamma$.
Using Eq.~(\ref{eq_GtCttau}) and $C(t) \equiv \Ctgam$, Eq.~(\ref{eq_CttauCtgam}) 
implies again Eq.~(\ref{eq_key}) \cite{foot_gamnot}.
\subsection{Oscillatory shear}
\label{theo_oscil}

Experimentally, the relaxation modulus $G(t)$ is, of course, commonly
sampled in a linear viscoelastic measurement \cite{WittenPincusBook,RubinsteinBook}
using an oscillatory shear imposing, e.g., a sinusoidal shear strain 
$\gamma(t) = \gamAmp \sin(\omega t)$ 
of amplitude $\gamAmp$ and angular frequency $\omega$ as shown in panel (d) of Fig.~\ref{fig_sketch}.
This implies an average shear stress
\begin{equation}
\tau(t) = \tauref + \gamAmp \left( \Gstor \sin(\omega t) + \Gloss \cos(\omega t) \right)
\label{eq_tauhat_oscil} 
\end{equation}
with $\tauref$ being the (not necessarily vanishing) reference shear stress at $\gamma=0$.
The storage modulus $\Gstor$ and the loss modulus $\Gloss$ may thus be determined 
as the Fourier coefficients 
\begin{eqnarray}
\Gstor & = & \frac{2}{p \Tperiod} \int_0^{p \Tperiod} \sin(\omega t) \frac{\tau(t)}{\gamAmp} \ddiff t\label{eq_Gstor_coeff} \\
\Gloss & = & \frac{2}{p \Tperiod} \int_0^{p \Tperiod} \cos(\omega t) \frac{\tau(t)}{\gamAmp} \ddiff t \label{eq_Gloss_coeff}
\end{eqnarray}
with $\Tperiod = 1/2\pi \omega$ being the period of the oscillation. 
Averages over instantaneous shear stresses $\tauhat(t)$ are performed over all periods $p$ sampled.
As stated, e.g., by Eq.~(7.149) and Eq.~(7.150) of Ref.~\cite{RubinsteinBook} both moduli are 
on the other side quite generally given by the Fourier-Sine and Fourier-Cosine transforms of $G(t)$ 
\begin{eqnarray}
\Gstor - \Geq & = & \omega \int_0^{\infty} (G(t)-\Geq) \sin( \omega t) \ddiff t,\label{eq_Gstor} \\
\Gloss        & = & \omega \int_0^{\infty} (G(t)-\Geq) \cos( \omega t) \ddiff t. \label{eq_Gloss} 
\end{eqnarray}
The latter two relations together with Eq.~(\ref{eq_key}) imply the key
relations Eq.~(\ref{eq_key_Gstor}) and Eq.~(\ref{eq_key_Gloss}) announced in the Introduction. 
We shall pay special attention to the low-$\omega$ and high-$\omega$ limits 
of $\Gstor$ and $\Gloss$ at the end of Sec.~\ref{simu_oscil}.
%

\begin{figure}[t]
\centerline{\resizebox{1.0\columnwidth}{!}{\includegraphics*{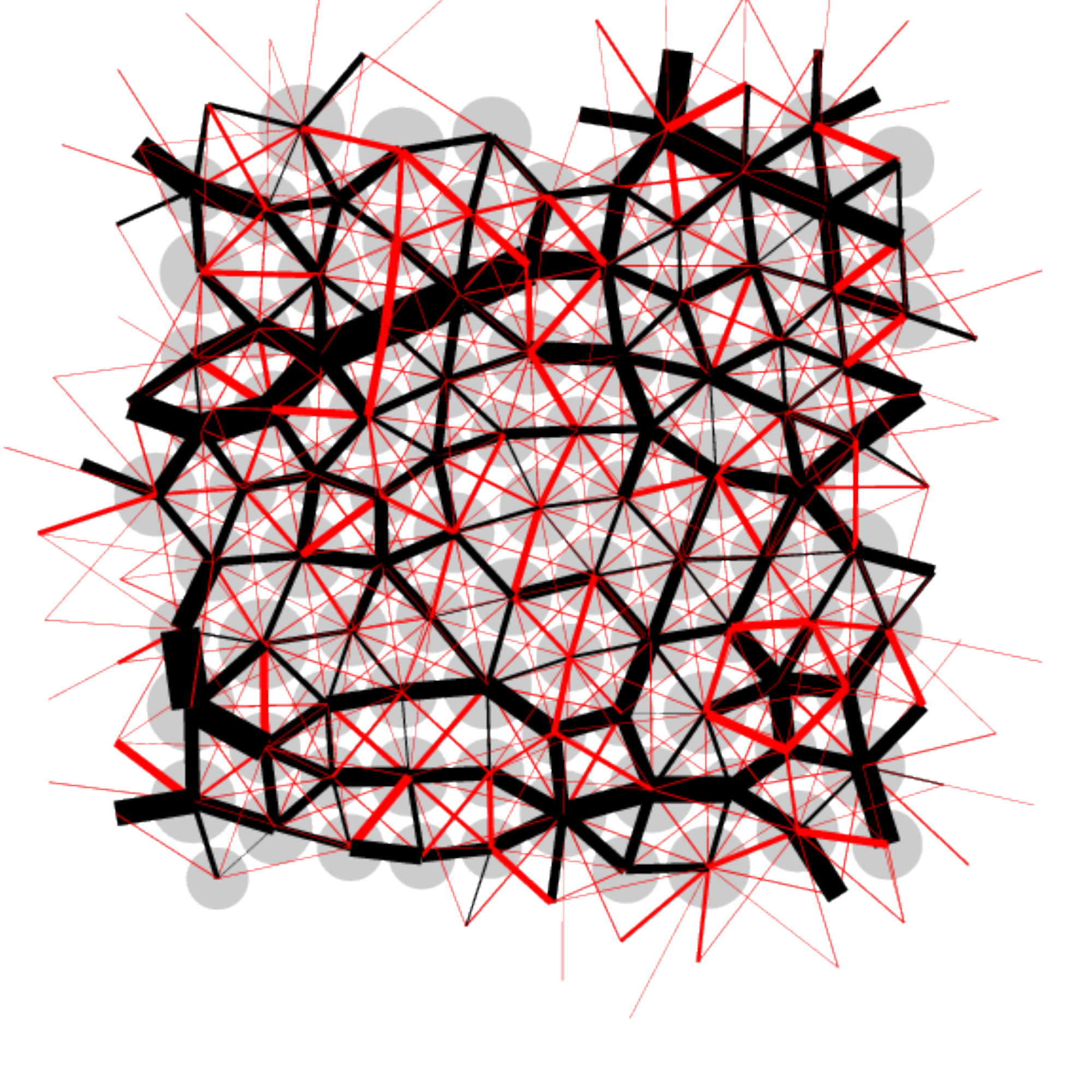}}}
\vspace*{-1.5cm}
\caption{Snapshot of a small subvolume of linear length $10$ containing about $100$ 
verticies of the elastic network considered in this work assuming
Eq.~(\ref{eq_Enet}). The total periodic box of linear length $L \approx 102.3$
contains $N=10^4$ verticies and $9956$ harmonic springs.
The light grey circles indicate the positions and diameters of the quenched 
polydisperse LJ bead system which was used for the construction of the
elastic network as described in Sec.~\ref{sec_refer}.
The lines represent the quenched forces of the athermal ($T=0$) reference configuration.
The dark (black) lines indicate repulsive forces between the verticies,
while the light (red) lines represent tensile forces.
The line width is proportional to the tension/repulsion.
\label{fig_network}
}
\end{figure}

\section{Model and algorithmic details}
\label{sec_model}

\paragraph*{Hamiltonian.}
To illustrate our key relations we present in Sec.~\ref{sec_simu} numerical data obtained 
using a periodic two-dimensional ($d=2$) network of harmonic springs.
The model Hamiltonian is given by the sum $\Hhat = \Hidhat + \Hexhat$ of 
a kinetic energy contribution $\Hidhat = m \sum_i \vvec_i^2 /2$ 
(assuming a monodisperse mass $m$) and an excess potential 
\begin{equation}
\Hexhat = \sum_l \ul(\rl) \mbox{ with } \ul(r) = \frac{1}{2} K_l \left(r - R_l\right)^2
\label{eq_Enet}
\end{equation}
where $K_l$ denotes the spring constant, $R_l$ the reference length and $\rl = |\rvec_i - \rvec_j|$ 
the length of spring $l$. The sum runs over all springs $l$ connecting pairs of beads $i$ and $j$ 
at positions $\rvec_i$ and $\rvec_j$. 
A small subvolume of the network considered is represented in Fig.~\ref{fig_network}.
An experimentally relevant example for such a permanent network is provided by 
endlinked or vulcanized polymer networks \cite{RubinsteinBook,DKG91,DKG94}.
Since the network topology is by construction {\em permanently fixed}, the shear response $G(t)$ 
must become finite for $t \to \infty$ for all temperatures at variance to systems with plastic 
rearrangements as considered, e.g., in Ref.~\cite{Biroli10}.
Note that the particle mass $m$ and Boltzmann's constant $\kB$ are set to unity. 
Lennard-Jones (LJ) units \cite{AllenTildesleyBook} are assumed throughout this paper.

\paragraph*{Computational methods, parameters and observables.}
The construction and the characterization of the reference network at zero temperature
is presented in Sec.~\ref{sec_refer}. As discussed in Sec.~\ref{sec_simu}, this network
is then investigated numerically by means of standard molecular dynamics (MD) simulation 
\cite{AllenTildesleyBook,FrenkelSmitBook} at constant particle number $N=10^4$, 
box volume $V \approx 102.3^2$ and a small, but finite mean temperature $T=0.001$.
Newton's equations are integrated using a velocity-Verlet algorithm with a tiny 
time step $\dtMD=10^{-4}$. The temperature $T$ is fixed using a Langevin thermostat 
with a friction constant $\zeta$ ranging from $\zeta=0.01$ up to $\zeta=5$ as specified below. 
An important property sampled in this study is the instantaneous stress tensor 
\cite{AllenTildesleyBook,foot_prime}
\begin{equation}
\sighat_{\alpha\beta} = -\frac{m}{V}\sum_{i=1}^N v_{i,\alpha} v_{i,\beta}
+ \frac{1}{V}\sum_{l} \rl \ul^{\prime}(\rl) \ n_{l,\alpha} n_{l,\beta} 
\label{eq_stresstens}
\end{equation}
where the Greek indices $\alpha$ and $\beta$ denote the $d=2$ spatial directions $x$ and $y$,
$v_{i,\alpha}$ the $\alpha$-component of the velocity of particle $i$
and $n_{l,\alpha}$ the $\alpha$-component of the normalized vector $\rvec_l$.
The first term in Eq.~(\ref{eq_stresstens}) gives the ideal gas contribution,
the second term corresponds to the Kirkwood excess stress for
pair interactions.
The trace of the stress tensor yields the instantaneous normal pressure 
$\hat{P} = - (\sighat_{xx}+\sighat_{yy})/d$,
the off-diagonal elements $\sighat_{xy} = \sighat_{yx}$ are consistent with the 
shear stress $\tauhat = \tauidhat + \tauexhat$, Eqs.~(\ref{eq_tauidhat},\ref{eq_tauexhat}), 
obtained in Sec.~\ref{theo_static}.
The shear strain $\gamma$ is set to zero for the computation of the autocorrelation function 
$C(t) = \beta V \left( \langle \tauhat(t)^2 \rangle - \langle \tauhat \rangle^2 \right)$.
A tiny step strain $\gamma \ll 1$ is imposed at $t=0$ in order to compute $G(t)$, 
as discussed in Sec.~\ref{simu_key}, and a sinusoidal strain $\gamma(t) = \gamAmp \sin(\omega t)$ 
to measure directly the storage and loss moduli $\Gstor$ and $\Gloss$ from the sine and cosine 
transforms of the instantaneous shear stress $\tauhat(t)$, Eqs.~(\ref{eq_Gstor_coeff},\ref{eq_Gloss_coeff}). 
As already described in the first paragraph of Sec.~\ref{theo_static}
we perform in both cases affine, canonical \cite{Goldstein} and (essentially) infinitessimal 
strain transformations of the box shape and the particle positions and momenta,
Eqs.~(\ref{eq_r_cantrans},\ref{eq_p_cantrans}). 
Note that the transformation of the momenta is, however, not important for the
simulations presented here where we focus on low temperatures and mainly on high values 
of the Langevin friction constant. Moreover, as shown in Ref.~\cite{WXB15}, 
similar results, especially concerning Eq.~(\ref{eq_key}),
are also obtained using Brownian dynamics or Monte Carlo simulations
\cite{AllenTildesleyBook,FrenkelSmitBook}.

\section{Reference configuration}
\label{sec_refer}

\paragraph*{Construction of network.}
As explained in detail in Ref.~\cite{WXP13}, our network has been constructed using the 
dynamical matrix $\Mmat$ of a polydisperse LJ bead glass comprising $N=10^4$ particles. 
Prior to forming the network a LJ bead system has been quenched to $T=0$ using a constant 
quenching rate and imposing a normal pressure $P=2$. 
The original LJ beads are represented in Fig.~\ref{fig_network} by grey polydisperse circles,
the permanent spring network created from the quenched bead system by lines between verticies.
The network has a number density $\rho \approx 0.96$ corresponding to a linear periodic box length $L \approx 102.3$.
Using Eq.~(\ref{eq_stresstens}) one determines a normal pressure $P \approx 2$ and a small, 
but finite shear stress $\tau \approx 0.011614$ (determined to high precision for reasons given below).
Using Eq.~(\ref{eq_muA}) one obtains the affine shear elasticity $\muA \approx 34.3$.
By construction the total force acting on each vertex of the reference network 
vanishes albeit the repulsive and tensile forces transmitted along each spring
do in general not as shown in Fig.~\ref{fig_network}. 
Due to the periodic boundary conditions the finite residual 
(normal and shear) stress does not relax.
Albeit being small, it is important to properly account for the finite, 
quenched shear stress $\tau$ of the reference for all the correlation functions considered.
We also note that the force network, Fig.~\ref{fig_network}, is strongly 
inhomogeneous with zones of weak attractive links embedded within a strong repulsive 
skeleton as discussed in Refs.~\cite{WTBL02,TWLB02}.

\begin{figure}[t]
\centerline{\resizebox{1.0\columnwidth}{!}{\includegraphics*{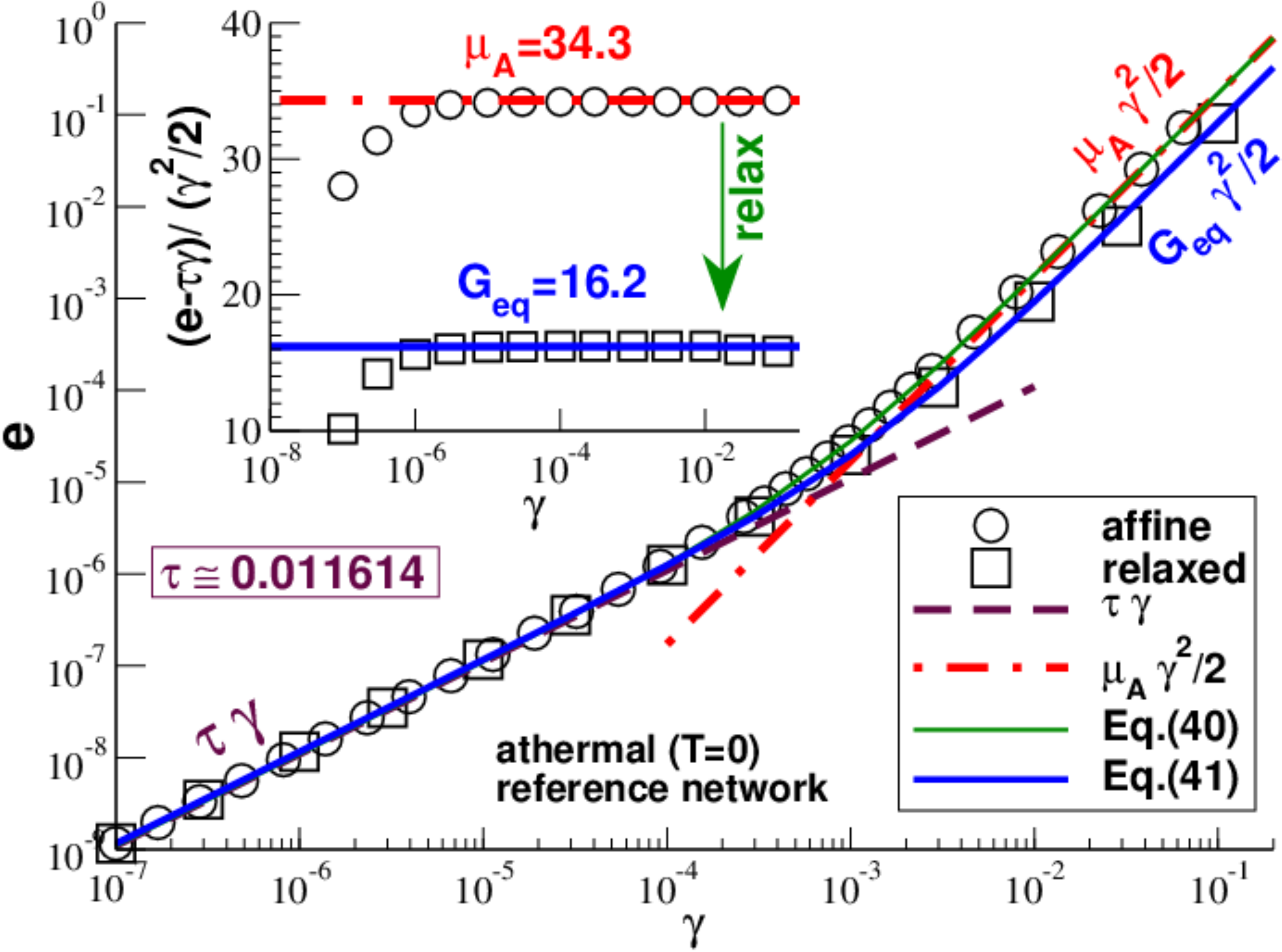}}}
\caption{Energy difference per volume $e$ after an affine strain $\gamma$ is applied (circles) 
and after the network has relaxed to the new ground state (squares). 
Main panel: Double-logarithmic representation of $e$ {\em vs.} $\gamma$.
The thin solid line indicates Eq.~(\ref{eq_Eaff}), the bold solid line Eq.~(\ref{eq_Enonaff2Geq}).
Inset: Half-logarithmic representation of reduced energy increment 
$(e - \tau \gamma) /(\gamma^2/2)$ {\em vs.} $\gamma$
where the residual shear stress $\tau = 0.011614$ has been taken off. 
The shear modulus $\Geq \approx 16.2$ is obtained from the rescaled 
groundstate energy (squares).
\label{fig_Egam}
}
\end{figure}

\paragraph*{Affine displacements.}
As shown in Fig.~\ref{fig_Egam}, it is possible to obtain numerically the shear stress $\tau$ 
and the affine shear elasticity $\muA$ from the energy increment per volume 
$e \equiv (\Hexhat(\gamma) - \Hexhat(0))/V$ caused by an affine shear strain $\gamma$ (spheres). 
In agreement with Eq.~(\ref{eq_Haffine}), the energy increases as
\begin{equation}
e = \tau \gamma + \muA \gamma^2/2
\label{eq_Eaff}
\end{equation}
as shown by the thin solid line in the main panel of Fig.~\ref{fig_Egam}.
The asymptotic limits for small and large $\gamma$ are indicated by, respectively,
the dashed line and the dash-dotted line.
We have used for the coefficients $\tau$ and $\muA$ in Eq.~(\ref{eq_Eaff})
the values obtained using Eq.~(\ref{eq_tauexhat}) and Eq.~(\ref{eq_muAexhat}).
This is merely a self-consistency check since both properties are actually {\em defined}
assuming a virtual {\em affine} strain transformation as reminded in Sec.~\ref{theo_static} 
\cite{WXP13,DoiEdwardsBook}. 
As shown in the inset, an accurate verification of $\muA$
over essentially the full $\gamma$-range is obtained by plotting in half-logarithmic coordinates
the reduced energy $(e - \tau \gamma)/(\gamma^2/2)$. 
For the smallest $\gamma$ an even more precise value of the substracted residual shear stress $\tau$
is required to collapse even these data points on the dash-dotted horizontal line.

\begin{figure}[t]
\centerline{\resizebox{1.0\columnwidth}{!}{\includegraphics*{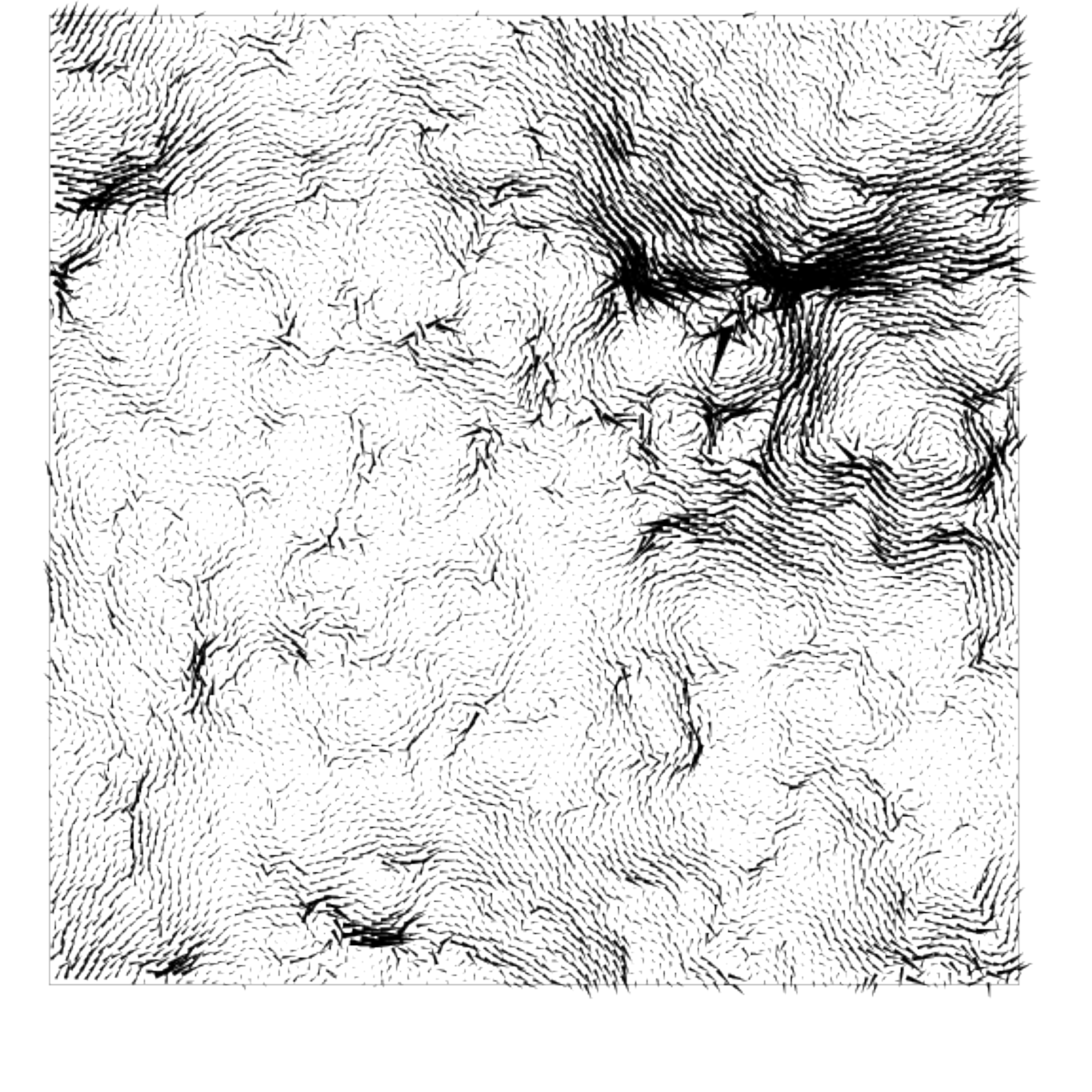}}}
\vspace*{-0.5cm}
\caption{Snapshot of non-affine displacement field after an affine shear strain with $\gamma=0.01$
is applied and the system is allowed to relax. 
The arrow length is proportionel to the non-affine particle displacements. 
The displacement field is correlated over distances much larger than the typical particle distance. 
All displacements are strictly linear in $\gamma$.
\label{fig_nonaffdisp}
}
\end{figure}

\paragraph*{Non-affine displacements.}
The forces $\fvec_i$ acting on the verticies $i$ of an affinely strained network 
do not vanish in general. Following Ref.~\cite{TWLB02} the network is first relaxed
by steepest descend, i.e. imposing displacements proportional to the force,
and then by means of the conjugate gradient method \cite{ThijssenBook}.
These non-affine displacements lower the final energy of the strained system as
indicated by the squares in Fig.~\ref{fig_Egam}. As shown by the bold solid line
in the main panel, these final energies scale as
\begin{equation}
e = \tau \gamma + \Geq \gamma^2/2. 
\label{eq_Enonaff2Geq}
\end{equation}
This scaling is similar to the affine strain energy, Eq.~(\ref{eq_Eaff}),
having the same linear term but with $\muA$ being replaced by the shear modulus $\Geq$. 
As shown in the inset, these energies can thus be used to determine $\Geq \approx 16.2$
by taking again into account the quenched shear stress at $\gamma=0$.
It follows from Eq.~(\ref{eq_Eaff}) and Eq.~(\ref{eq_Enonaff2Geq})
that due to the non-affine displacements a substantial fraction
$1-\Geq/\muA \approx 1/2$ of the affine strain energy is relaxed 
for large $\gamma$ in agreement with Ref.~\cite{TWLB02}.
A snapshot of these non-affine displacements is given in Fig.~\ref{fig_nonaffdisp}.
Note that the non-affine displacements are correlated over distances much larger 
than the typical particle distance and one thus expects deviations from 
standard continuum mechanics if similar length scales are probed.

\begin{figure}[t]
\centerline{\resizebox{1.0\columnwidth}{!}{\includegraphics*{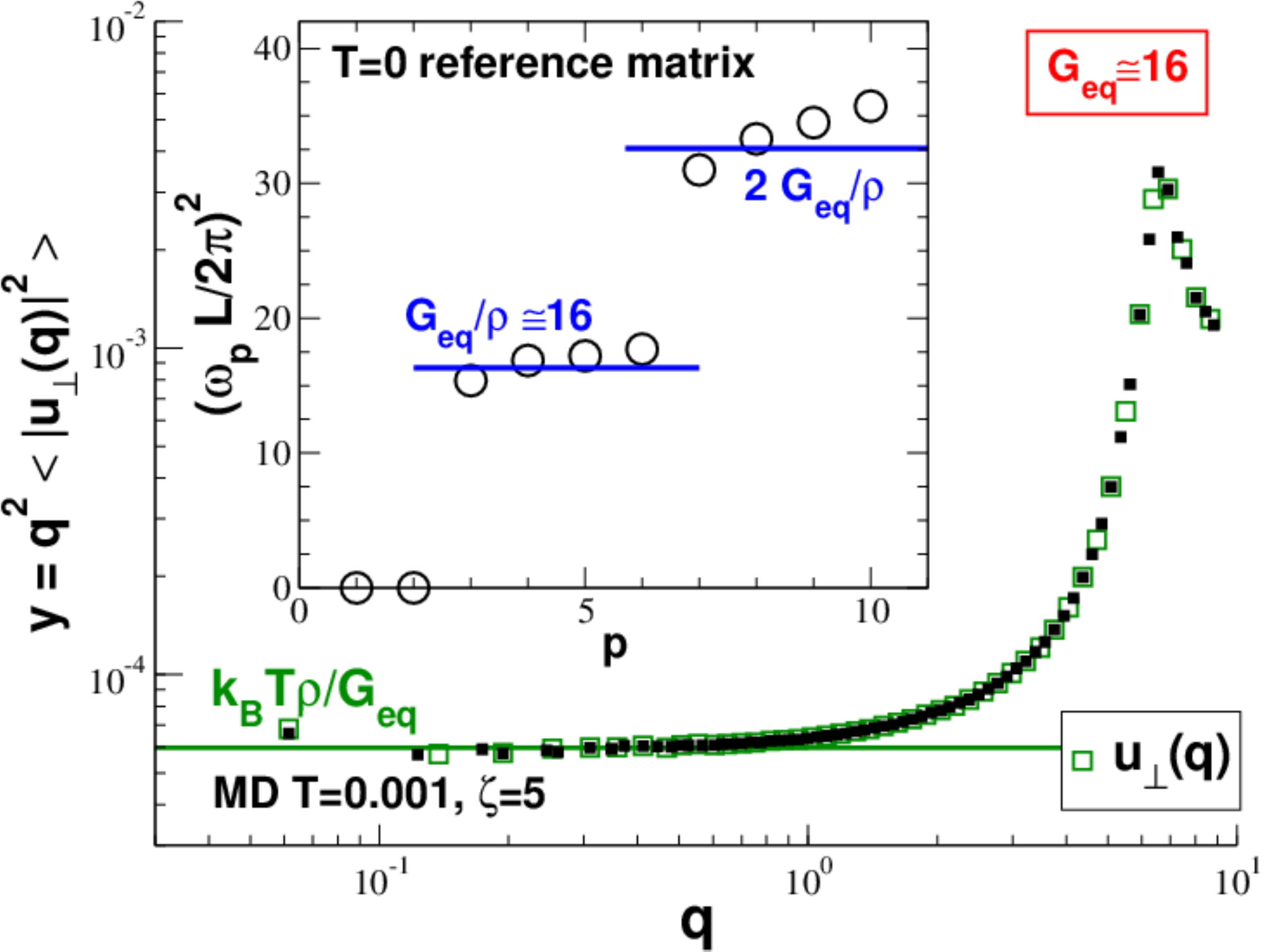}}}
\caption{Determination of shear modulus $\Geq$.
Inset: Lowest eigenfrequencies $\omega_p$ obtained by diagonalization of the
dynamical matrix. The bold horizontal lines correspond to $\Geq/\rho \approx 16$ 
according to Eq.~(\ref{eq_omegap}) fitting the eigenvalues for $p=3,4,5,6$.
Main panel: Fourier transformed transverse displacement field for a temperature $T=0.001$ 
as a function of wavevector $q$. The horizontal line corresponds again to $\Geq \approx 16$.
The open symbols indicate data obtained using the ground state positions as reference, 
the filled symbols assume the average vertex position as reference.
\label{fig_static}
}
\end{figure}

\begin{figure}[t]
\centerline{\resizebox{1.0\columnwidth}{!}{\includegraphics*{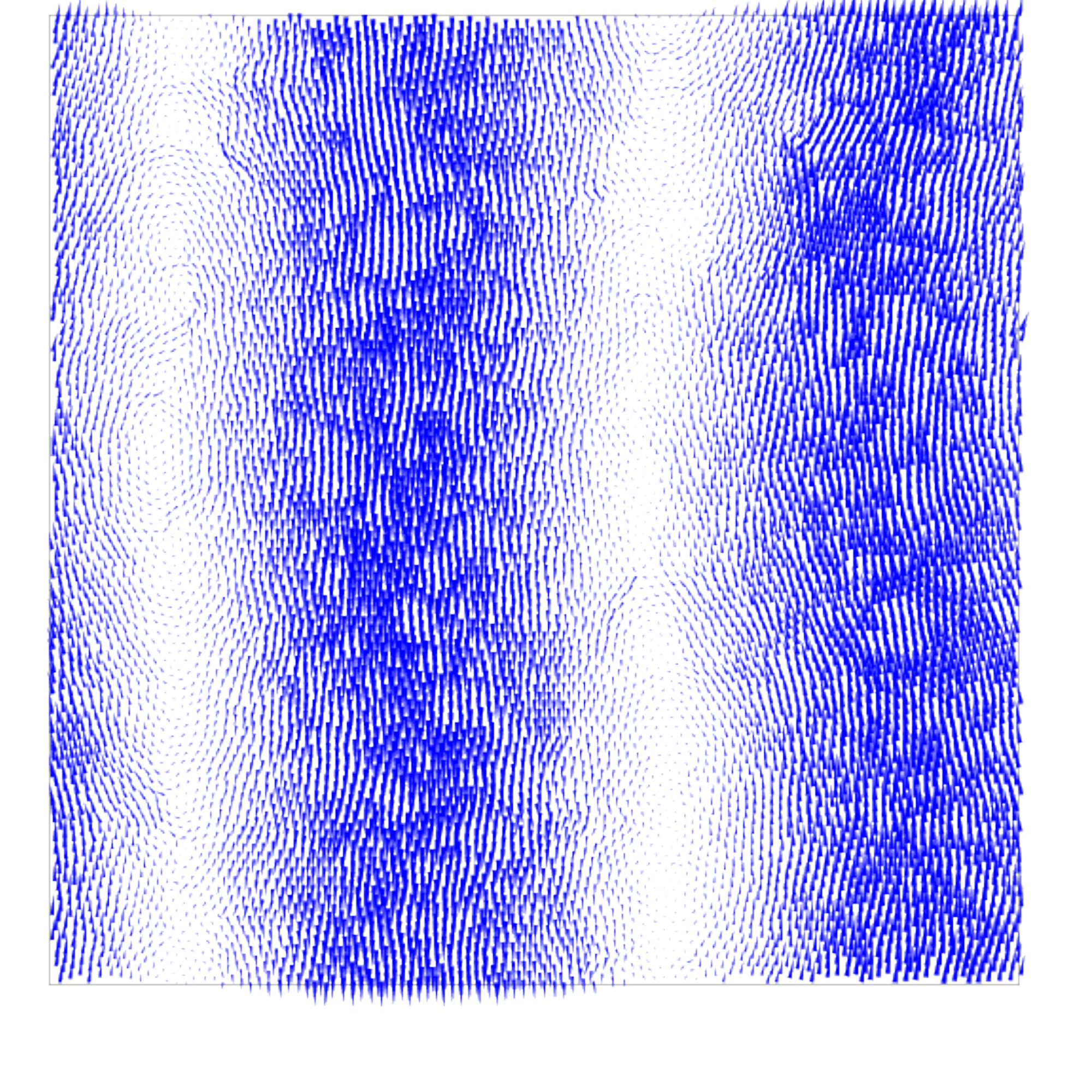}}}
\vspace*{-1.0cm}
\caption{Eigenvector for $p=4$ ($n=1,m=0$) corresponding to a transverse 
planewave with wavector $\qvec$ pointing in the horizontal direction,
i.e. with displacements essentially in the vertical direction, and wavelength $\lambda=L$.
Interestingly, heterogeneous deviations from the planewave solution
of continuum mechanics are even visible for this low eigenmode.
\label{fig_vect4}
}
\end{figure}

\paragraph*{Eigenstates of the reference.}
Following Refs.~\cite{WTBL02,TWLB02} the shear modulus $\Geq$ of the network 
at $T=0$ may alternatively be computed from the lowest non-trivial eigenfrequencies $\omega_p$.
(The running index $p$ increases with frequency.) This is shown in the inset of Fig.~\ref{fig_static}. 
The eigenfrequencies $\omega_p^2$ have been determined by diagonalization of the 
dynamical matrix $\Mmat$ by means of Lanczos' method \cite{ThijssenBook}.  
It follows from continuum elasticity \cite{LandauElasticity} that the eigenfunctions 
must be planewaves with wavevectors $\qvec$ quantified by the boundary conditions, 
i.e. $(q_x,q_y) = (2\pi/L) \ (n,m)$ with $n,m=0, \pm 1,\ldots$ being two quantum numbers.
An example for such a planewave is given in Fig.~\ref{fig_vect4} for $p=4$ and 
the pair of quantum numbers $(n,m)=(1,0)$.
For the wavelength this implies
\begin{equation}
\lambda_p = 2\pi / |\qvec| = L / \sqrt{n^2+m^2}
\label{eq_lambdap}
\end{equation}
and for the eigenfrequencies of the transverse modes
\begin{equation}
\omega_p = \frac{2\pi \ctrans}{L} \sqrt{n^2 + m^2} \mbox{ with } \ctrans = \sqrt{\Geq/\rho}
\label{eq_omegap}
\end{equation}
being the transverse wavevelocity.
As shown by the horizontal lines, we obtain by fitting Eq.~(\ref{eq_omegap}) 
to the frequencies for $p=3,4,5,6$ (corresponding to $n^2+m^2=1$) that 
$\Geq \approx 16$ and $\ctrans \approx 4$. Interestingly, the degeneracy of
the eigenvalues expected from continuity elasticity is already lifted for
$p=7,8,9,10$ ($n^2+m^2=2$), i.e. the box size $L$ does not allow a
precise determination of $\Geq$ using these eigenvalues.
Deviations from the planewave solution are even visible from
the eigenvector for $p=4$ presented in Fig.~\ref{fig_vect4}.
Continuum elasticity must break down in any case if wavelengths of order of about the
particle distance are probed, $\lambda \approx 1$. This implies that
Eq.~(\ref{eq_omegap}) can at best hold up to a frequency 
$\omega = 2\pi \ctrans/\lambda \approx 25$.
We come back to this issue in Sec.~\ref{simu_static} and Sec.~\ref{simu_oscil}.
%

\section{Finite-temperature computational results}
\label{sec_simu}

\subsection{Some static properties}
\label{simu_static}

\paragraph*{Introduction.}
Since the temperature $T = 0.001$ is rather small, one expects all static properties such as  
the pressure $P$ or the elastic modulus $\Geq$ to be similar to their ground state values.
As we have checked comparing various methods one confirms indeed that 
$P \approx \Pex \approx 2$, $\tau \approx \tauex \approx 0.011$,
$\muA \approx \muAex \approx 34$, $\muF \approx \muFex \approx 18$ and $\Geq \approx 16$.
The same applies in fact to all small temperatures $T \ll 1$.

\paragraph*{Displacement correlations.}
A finite-$T$ method for computing $\Geq$, which is also of experimental relevance \cite{Klix12},
is presented in the main panel of Fig.~\ref{fig_static}.
Using an ensemble of $10^4$ configurations sampled over a total time interval $10^6$
with $\zeta=5$ we obtain first the displacements $\uvec_i$ for each vertex particle $i$ of the
network either by taking the position of the ground state network as reference
for defining the displacement (open squares) or alternatively the average 
particle position in the ensemble (filled squares). 
The displacement field $\uvec(\rvec) = \sum_i \uvec_i \delta(\rvec - \rvec_i)$
is then Fourier transformed according to
$\uvec(\qvec) = \sum_{i=1}^N \exp(i \qvec \cdot \rvec) \ \uvec(\rvec) /\sqrt{N}$ \cite{Klix12}.
Note that the wavevector $\qvec$ must be commensurate to the square simulation
box of linear length $L$, Eq.~(\ref{eq_lambdap}). The component of $\uvec(\qvec)$ 
perpendicular to $\qvec$
corresponds to the transverse component $\utrans(\qvec)$ of the Fourier transformed 
displacement field. Using that according to the equipartition theorem every independent
elastic mode corresponds to an average kinetic or potential energy $\kBT/2$, 
continuum mechanics implies that \cite{LandauElasticity,Klix12}
\begin{equation}
y \equiv q^2 \la | \utrans(\qvec) |^2 \ra \to \kBT \rho/\Geq \ \mbox{ for } q \to 0
\label{eq_utrans2Geq}
\end{equation}
where the average $\langle \ldots \rangle$ is taken over the ensemble of stored configurations.
As can be seen from Fig.~\ref{fig_static}, $y$ becomes rapidly constant below $q \approx 1$
confirming $\Geq \approx 16$ as indicated by the bold horizontal line.
Incidentally, both definitions of the displacement field yield identical results.
Since only the second definition using the average particle position can normally
be used in experimental studies, this is rather reassuring. Note also that
the first peak for large wavevectors corresponds to a wavelength
$\lambda = 2\pi/q \approx 1$ of order of the typical particle distance
where continuum mechanics should break down.

\begin{figure}[t]
\centerline{\resizebox{1.0\columnwidth}{!}{\includegraphics*{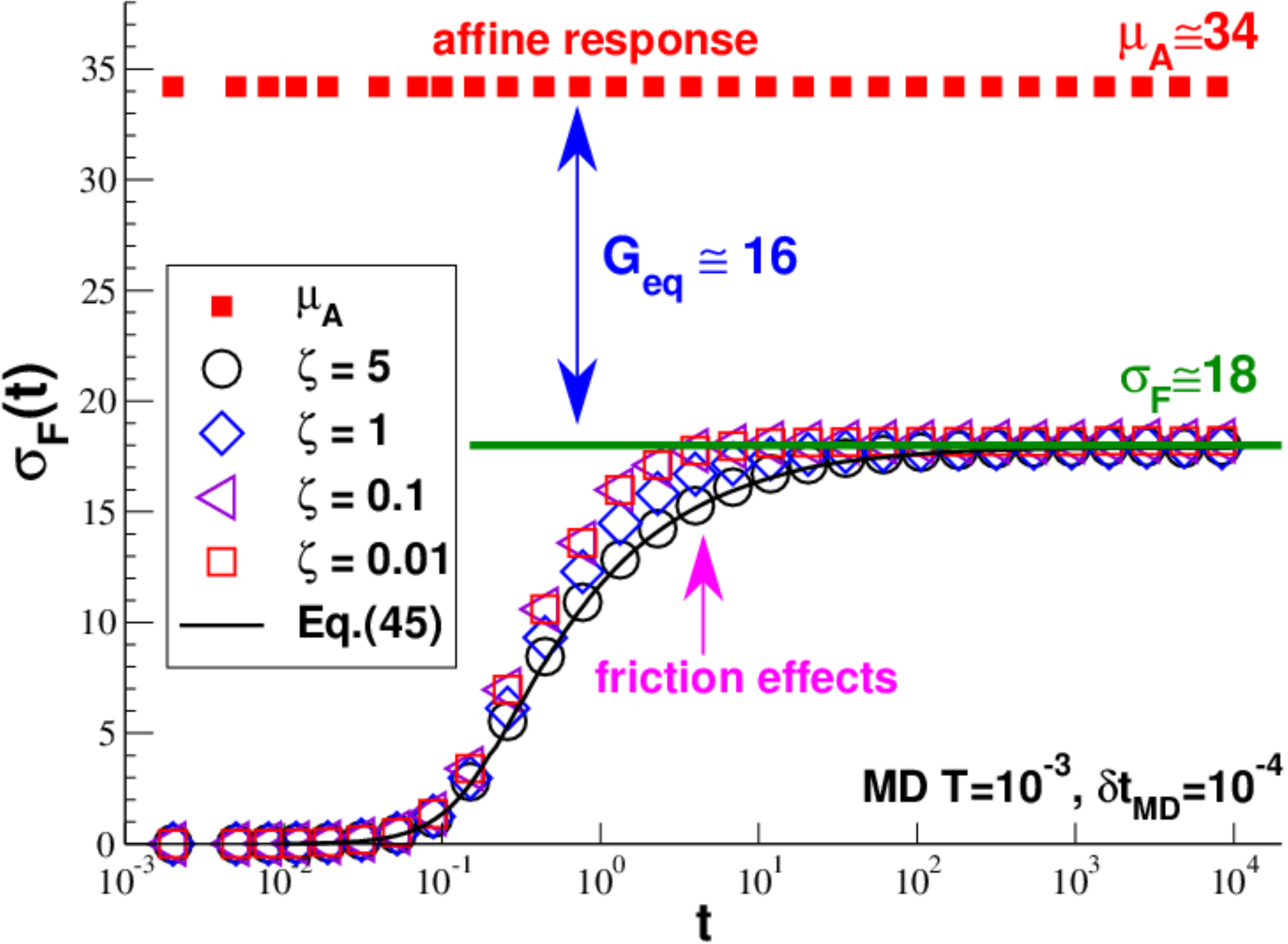}}}
\caption{Determination of $\Geq \approx 16$ using Eq.~(\ref{eq_GeqNVgT}) 
as a function of the sampling time $t$ at $T=0.001$.
While the simple average $\muA \approx 34$ converges immediately (small filled squares),
the stress fluctuations $\muF(t)$ are seen to increase monotonously 
to the large-$t$ limit $\muF \approx 18$ (bold solid line).
This is shown for four different friction constants $\zeta = 5, 1, 0.1$ and $0.01$. 
The thin solid line compares the data for $\zeta=5$ with the integrated 
correlation function $C(t)$ using Eq.~(\ref{eq_muFtCt}). 
\label{fig_muFt}
}
\end{figure}

\paragraph*{Stress fluctuations.}
While it is thus possible to get $\Geq$ from the displacement correlations,
it is from the computational point of view more convenient to determine the
modulus using the stress-fluctuation formula, Eq.~(\ref{eq_GeqNVgT}).
This is shown in Fig.~\ref{fig_muFt} where the affine shear elasticity $\muA$ (small filled squares)
and the stress-fluctuation term $\muF(t)$ are presented as functions of the sampling time $t$.
(The notation $\muF$ without time argument refers to the static thermodynamic large-$t$ limit,
while $\muF(t)$ indicates that this property has been determined using a finite time window $t$.)
While the simple average $\muA$ is obtained immediately, $\muF(t)$ 
is seen to increase monotonously from zero to the large-$t$ plateau $\muF \approx 18$
(horizontal bold solid line).  
Friction effects are important only in the intermediate sampling time window between 
$t \approx 0.1$ and $10$ where the stress fluctuations strongly increase.
The data obtained for Langevin friction constants $\zeta = 0.1$ and $0.01$ are essentially identical.
The time dependence of $\muF(t)$ for different friction constants is further analysed at 
the end of Sec.~\ref{simu_key}.
\begin{figure}[t]
\centerline{\resizebox{1.0\columnwidth}{!}{\includegraphics*{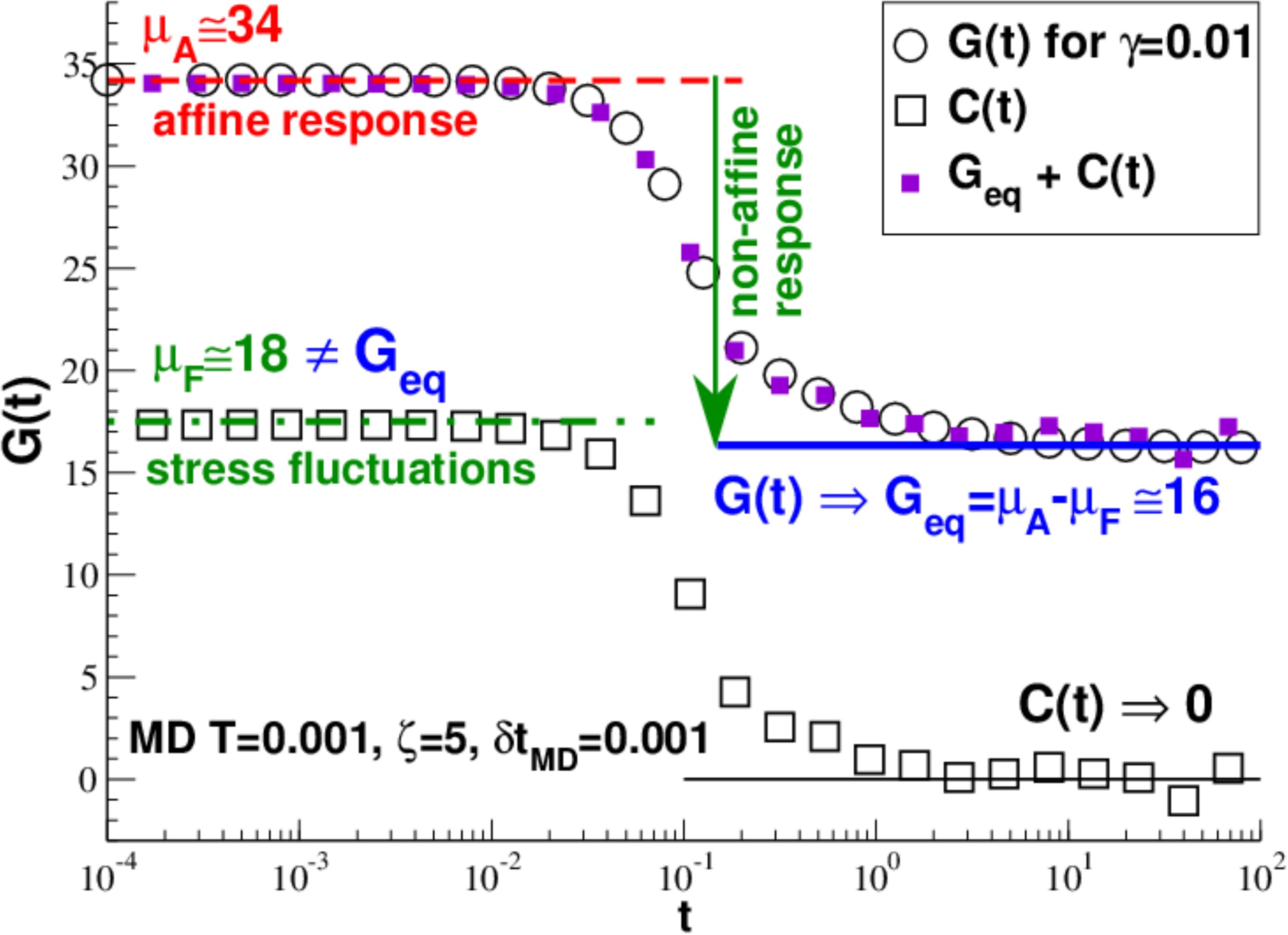}}}
\caption{Stress relaxation modulus $G(t)$ and autocorrelation function $C(t)$
sampled by means of a Langevin thermostat of friction constant $\zeta=5$.
An affine step strain $\gamma=0.01$ is applied at $t=0$ to determine $G(t)$.
\label{fig_keyA}
}
\end{figure}

\subsection{Time domain of key relations} 
\label{simu_key}

\paragraph*{High friction limit.}
We turn now to the numerical verification of the key relations stated in the Introduction
focusing first on results obtained in the time domain which have already been reported in Ref.~\cite{WXB15}.
The data presented in Fig.~\ref{fig_keyA} have been computed using a Langevin thermostat 
with one high friction constant $\zeta=5$. 
This is done merely for presentational reasons since in this limit the 
kinetic degrees of freedom can be disregarded and oscillations are suppressed.
The stress relaxation modulus $G(t)$ given has been computed as indicated in panel (b)
of Fig.~\ref{fig_sketch} by applying an affine canonical shear strain $\gamma=0.01$ 
and by averaging over 1000 runs starting from independent configurations at $t=0^{-}$.
Due to the strong damping $G(t)$ decreases monotonously from $G(0^{+})=\muA$ to $\Geq$ 
while $C(t)$ decays from $C(0) = \muF$ to zero.
Confirming Eq.~(\ref{eq_key}), only after vertically shifting 
$C(t) \rightarrow C(t)+\Geq$ one obtains a collapse on the directly computed modulus $G(t)$.
This is the most important numerical result of the present work.

\begin{figure}[t]
\centerline{\resizebox{1.0\columnwidth}{!}{\includegraphics*{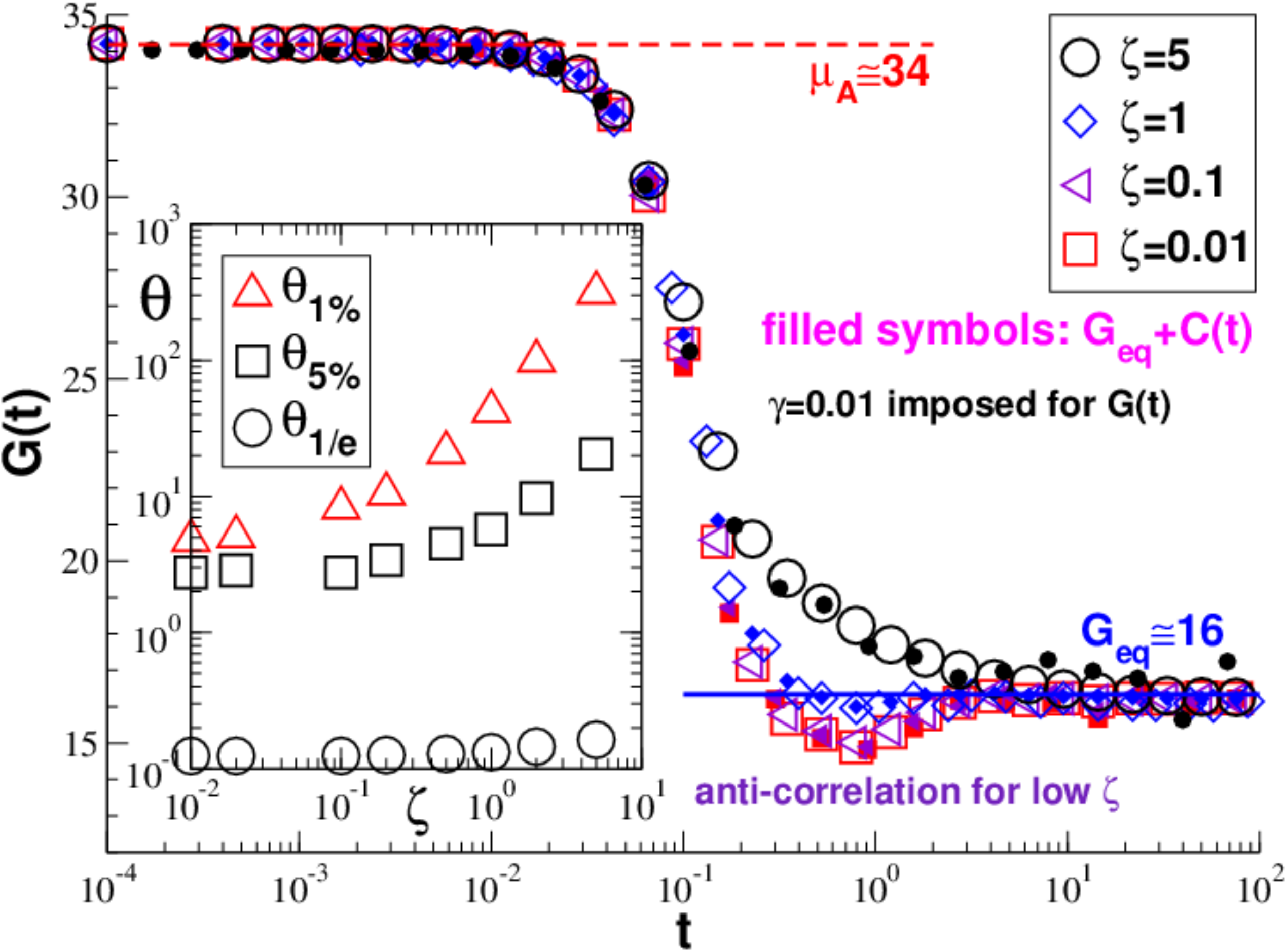}}}
\caption{Stress relaxation modulus $G(t)$ obtained using a step strain (open symbols) and 
the rescaled autocorrelation function $\Geq + C(t)$ at $\gamma=0$ (filled symbols) 
for several friction constants $\zeta$.
Inset: 
Shear stress relaxation time $\theta(\zeta)$ measured using 
$\thetae$, $\thetafif$ and $\thetaone$ defined in the main text.
\label{fig_keyB}
}
\end{figure}

\paragraph*{Friction effects and stress relaxation time.}
A similar scaling collapse of $G(t)$ and $C(t)+\Geq$ has been also obtained for different 
friction constants $\zeta$ as may be seen from 
Fig.~\ref{fig_keyB}. As one expects, the MD data decays more rapidly with decreasing $\zeta$ 
and reveals anti-correlations for the lowest $\zeta$ probed. 
The effect of the friction constant may be compactly
described using the stress relaxation time $\theta$ shown in the inset of Fig.~\ref{fig_keyB}.
We compare here three different definitions. 
The definition $C(t=\thetae)=\muF/e$, indicated by spheres,
describes the time where the correlation functions first becomes small. 
This simple definition underestimates correlations at larger times.
Alternatively, one may attempt to define $\theta$ by the ratio $c_1/c_0$ of the first 
to the zeroth moment of $C(t)$ with $c_n \equiv \int_0^{\infty} t^n C(t) \ddiff t$
(not shown). 
Unfortunately, $c_1$ is found to converge badly and we have not been able to 
obtain reliable values over the full $\zeta$-range. 
A numerically well-behaved alternative is based on 
an exact relation between the autocorrelation function $C(t)$ and the 
monotoneously increasing fluctuation term $\muF(t)$ determined as a 
function of the sampling time $t$.
Assuming time-translational invariance it can be shown \cite{WXB15} that
\begin{equation}
y(t) \equiv 1-\frac{\muF(t)}{\muF} = \frac{2}{t} \int_0^t \ddiff s \ (1 - s/t) \ \frac{C(s)}{\muF},
\label{eq_muFtCt}
\end{equation}
i.e. the dimensionless ratio $y(t)$ characterizes the {\em difference} 
of the zeroth and the first moment of the autocorrelation function integrated up to $t$.
That this relation holds can be seen from the thin line presented in Fig.~\ref{fig_muFt}.
We note {\em en passant} that $y(t)$ must ultimately vanish slowly as $1/t$ since
the integral over $C(s)$ in Eq.~(\ref{eq_muFtCt}) becomes constant \cite{WXB15}.
(This $1/t$-decay is consistent with the general finite-sampling time 
corrections for fluctuations \cite{LandauBinderBook}.)
Since $\muF(t)$ can be determined directly,
one may define $\theta(\zeta)$ using a fixed ratio $y(\theta)$.
We have presented in Fig.~\ref{fig_keyB} the constants $y(\thetafif) = 5 \%$ (squares) 
and $y(\thetaone) = 1 \%$ (triangles). Note that $\thetafif$ corresponds to
the time where $\muF(t)$ begins to saturate, while $\thetaone$ indicates the time
where the stress-correlations become neglible and $\muF(t)$ thus allows
a good estimation of $\Geq$ using the stress-fluctuation formula.

\subsection{Oscillatory shear}
\label{simu_oscil}

\begin{figure}[t]
\centerline{\resizebox{1.0\columnwidth}{!}{\includegraphics*{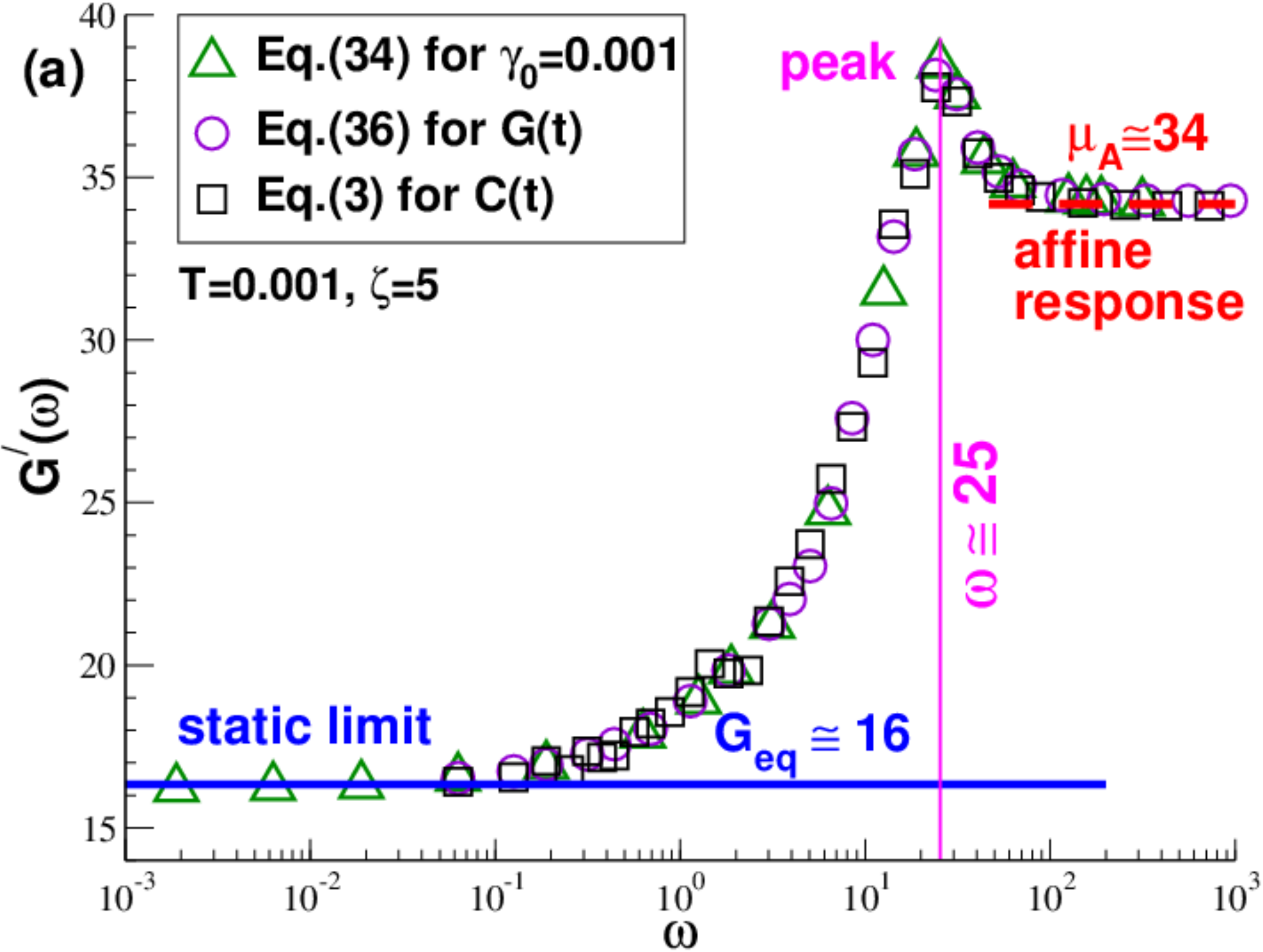}}}
\centerline{\resizebox{1.0\columnwidth}{!}{\includegraphics*{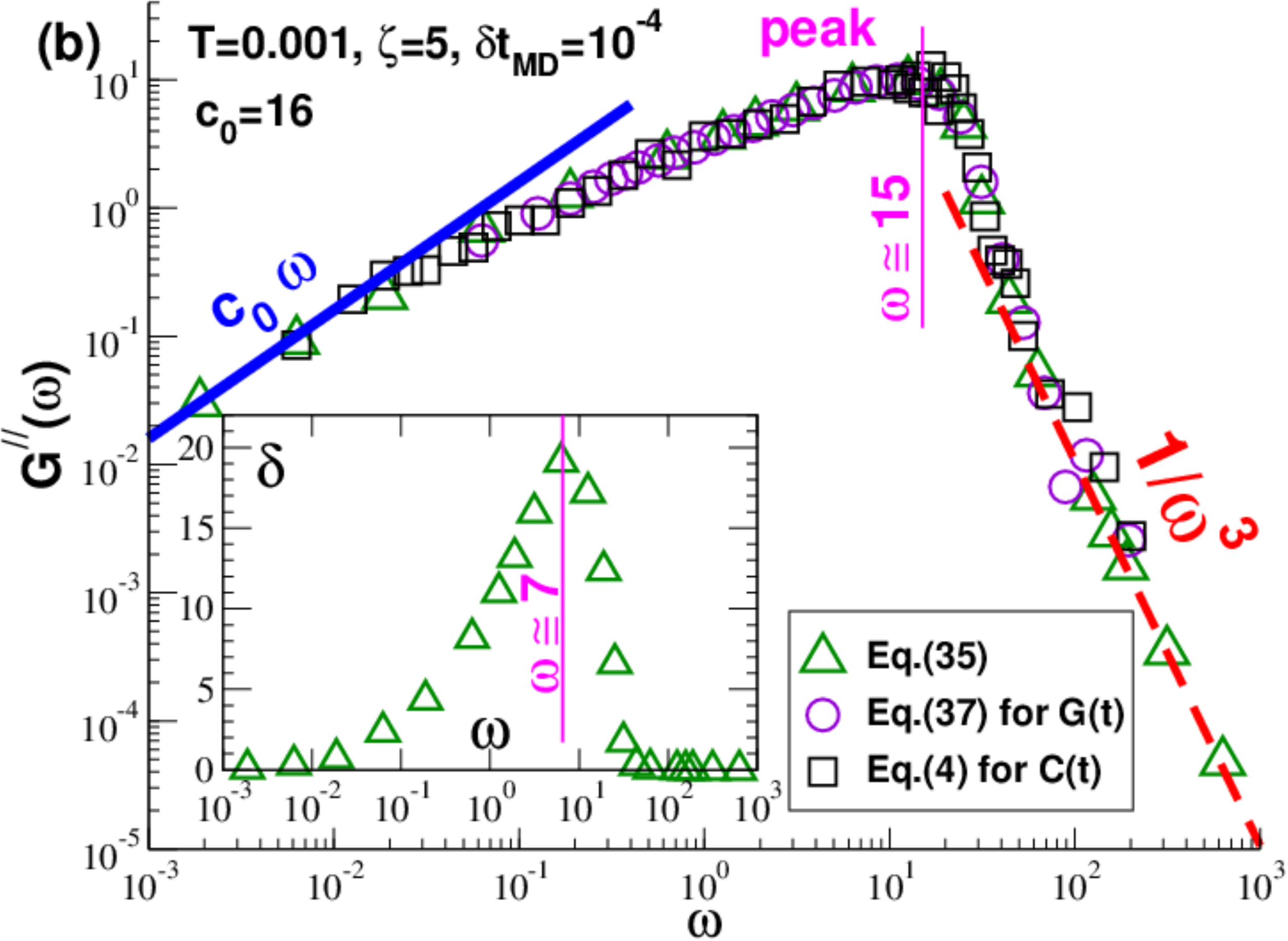}}}
\caption{Oscillatory shear moduli 
{\bf (a)} $\Gstor$ using a log-linear representation and
{\bf (b)} $\Gloss$ using a double-logarithmic representation 
for a large friction constant $\zeta=5$.
We compare the Fourier coefficients (triangles) obtained by applying a sinusoidal 
shear strain with amplitude $\gamAmp=0.001$ to the Fourier transformed step-strain 
relaxation modulus $G(t)$ (circles) and shear autocorrelation function $C(t)$ (squares). 
The bold solid lines indicate the expected asymptotic limits for small $\omega$,
the dashed lines the large-$\omega$ asymptotics, the thin vertical lines the
frequencies $\omega \approx 25$ and $\omega \approx 15$ of the maxima of both moduli.
The latter values coincide with the breakdown of continuum elasticity 
at a planewave frequency $\omega = 2\pi \ctrans/\lambda$ with $\lambda \approx 1$
probing the typical particle distance. 
Inset of {\bf (b)}: Phase angle $\delta$ with maximum  $\delta \approx 20^0$ 
at $\omega \approx 7$ (vertical line).
\label{fig_GstorGloss}
}
\end{figure}

\begin{figure}[t]
\centerline{\resizebox{1.0\columnwidth}{!}{\includegraphics*{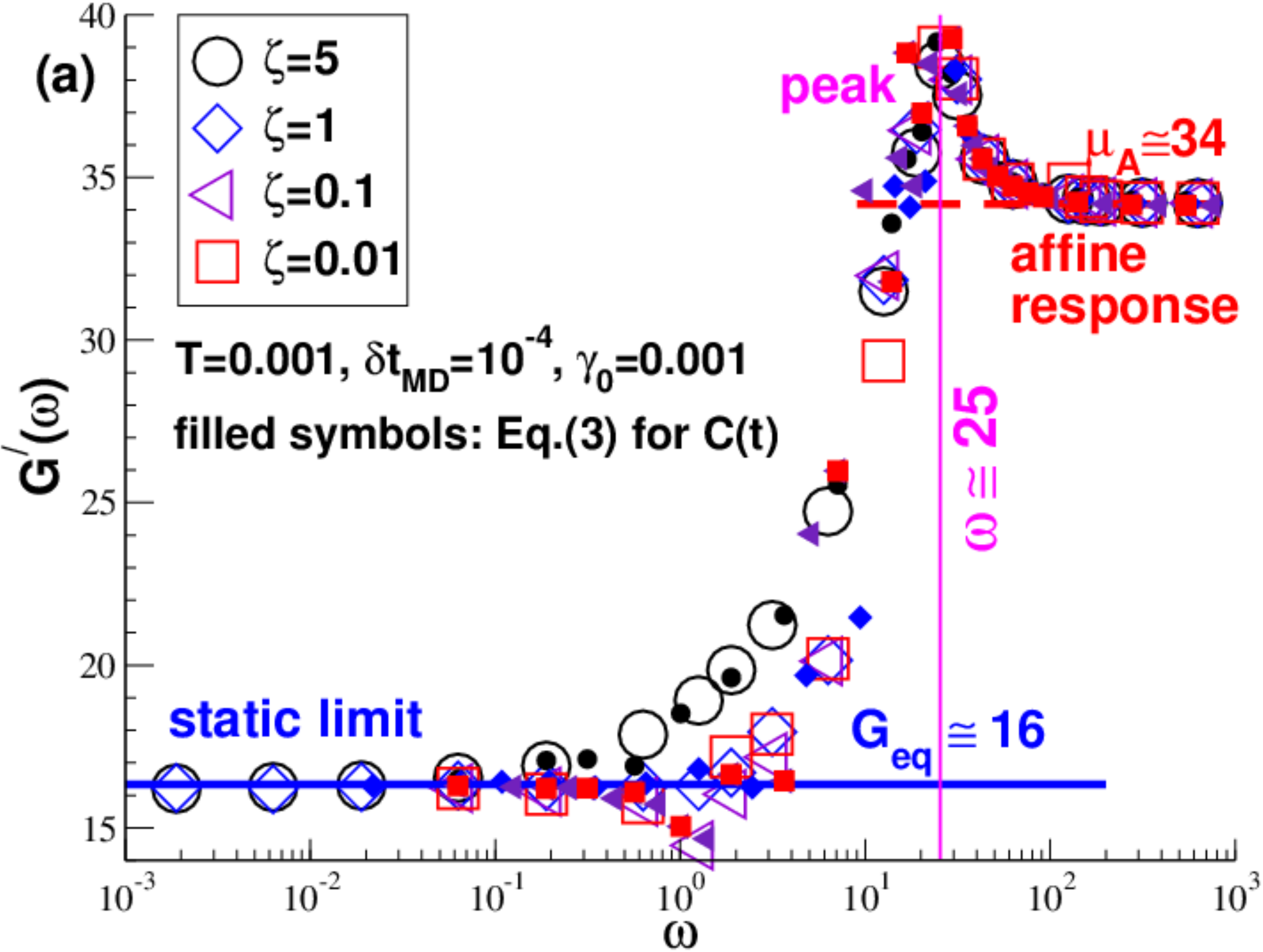}}}
\centerline{\resizebox{1.0\columnwidth}{!}{\includegraphics*{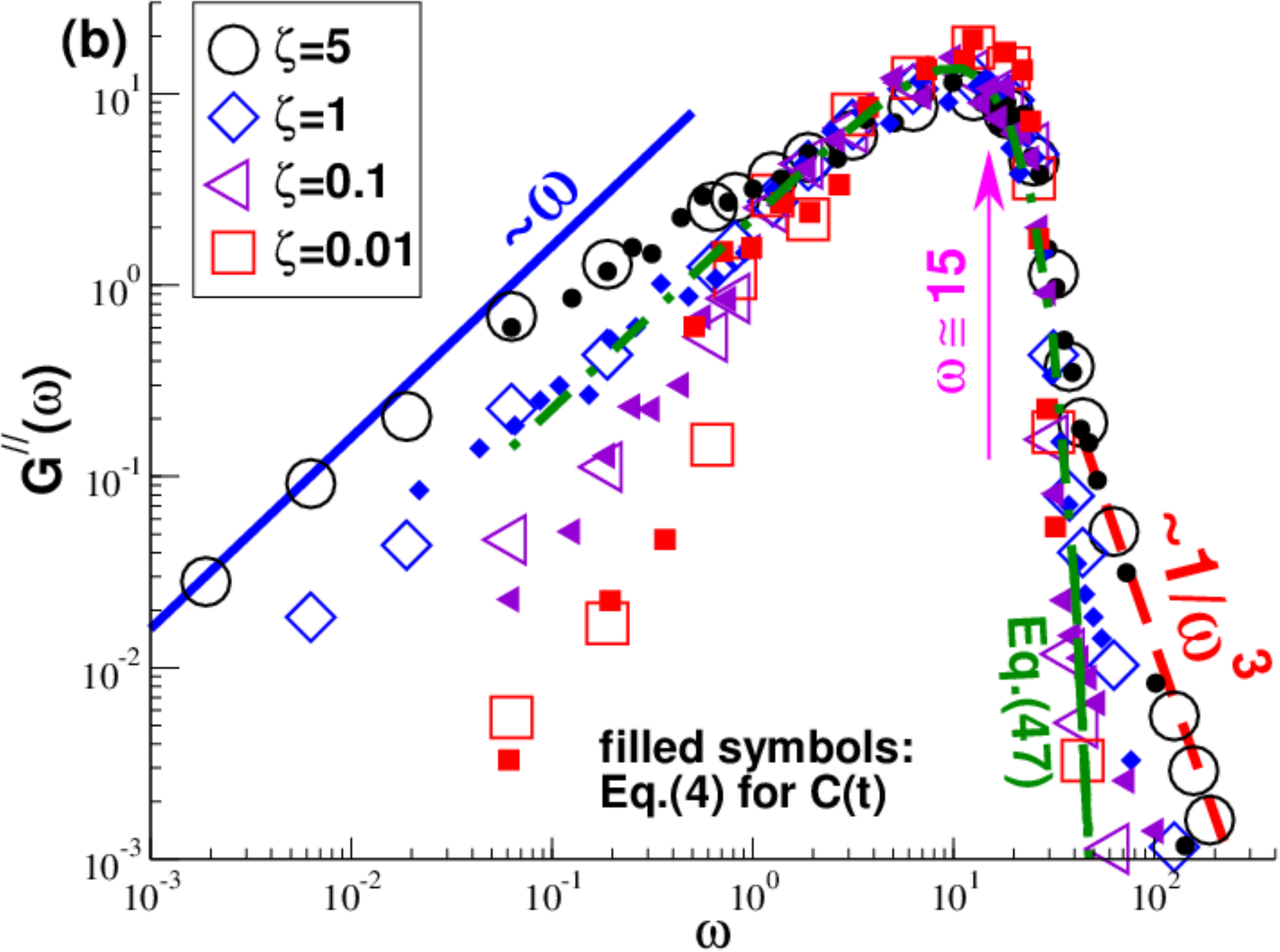}}}
\caption{Effect of Langevin friction constant $\zeta$ with open symbols referring to
data obtained using oscillatory shear with amplitude $\gamAmp=0.001$:
{\bf (a)} Storage modulus $\Gstor$. 
The friction constant $\zeta$ only matters in an intermediate frequency range between 
$\omega \approx 0.1$ and $\omega \approx 10$ below the peak at $\omega \approx 25$.
{\bf (b)} Loss modulus $\Gloss$. 
The dot-dashed line is a phenomenological fit for large $\omega$ and small $\zeta$
assuming Eq.~(\ref{eq_Gaussfit}).  
\label{fig_GstorGloss_zeta}
}
\end{figure}

\paragraph*{Direct determination of $\Gstor$ and $\Gloss$.}
As already stressed in Sec.~\ref{theo_oscil}, the relaxation modulus $G(t)$ is
commonly determined experimentally by inverse Fourier transformation of the storage
modulus $\Gstor$ and/or the loss modulus $\Gloss$ obtained in a linear viscoelastic 
measurement imposing an oscillatory shear \cite{WittenPincusBook,RubinsteinBook}. 
Motivated by this we perform non-equilibrium MD simulations in the linear response limit 
by imposing a sinusoidal shear strain of frequency $\omega$ with an amplitude $\gamAmp=0.001$.
(By varying $\gamAmp$ it has been checked that all reported values are in the linear regime.)
This is done by performing every time step $\dtMD=10^{-4}$ an affine canonical strain,
Sec.~\ref{theo_static}. 
We use production runs of total length $p \Tperiod$ with $\Tperiod = 2\pi/\omega$ and 
at least $p=100$ oscillation periods, i.e. the computational load increases as 
$p/\omega$ with decreasing frequency.
This sets a lower limit $\omega \approx 10^{-3}$ for the angular frequencies we have been able to sample.
Two production runs are compared to rule out transient behavior.
The instantaneous shear stress $\tauhat(t)$ is sampled each $\dtMD$.
Using Eq.~(\ref{eq_Gstor_coeff}) and Eq.~(\ref{eq_Gloss_coeff}) one obtains $\Gstor$ and $\Gloss$
as indicated by open triangles in Fig.~\ref{fig_GstorGloss},
where we focus for simplicity on one value $\zeta=5$ of the friction constant,
and large open symbols in Fig.~\ref{fig_GstorGloss_zeta},
where data for different $\zeta$ are presented. 

\paragraph*{Test of key relations.}
These values of $\Gstor$ and $\Gloss$ are compared in Fig.~\ref{fig_GstorGloss}
with the Sine-Fourier and Cosine-Fourier transforms of the shear modulus $G(t)$ obtained from the 
step strain experiment (circles) and of the shear stress autocorrelation function $C(t)$.
To reduce the well-known (but considering our tiny time step $\dtMD$ surprisingly severe) 
numerical problems at large frequencies ($\omega \gg 10$), 
Filon's quadrature method \cite{AllenTildesleyBook} is used for the Fourier integration.
These Fourier transforms collapse nicely on the respective directly computed moduli $\Gstor$ and $\Gloss$. 
The crucial point is here that it is $\Geq + \Cstor$ which collapses on $\Gstor$, 
while $\Cstor$ does {\em not} being much too small considering that $\Geq \approx 16$, as seen in panel (a).
We have verified that a similar data collapse for $\Geq + \Cstor$ onto $\Gstor$ and $\Closs$ onto $\Gloss$
is also obtained for other friction constants as may be seen from Fig.~\ref{fig_GstorGloss_zeta}.
Having settled this fundamental scaling issue let us turn finally to the description of
the general shape and asymptotic behavior of both moduli.

\paragraph*{Low-frequency limit.}
The low-$\omega$ limit of $\Gstor$ and $\Gloss$ are readily obtained by expanding 
$\sin(x)$ in Eq.~(\ref{eq_key_Gstor}) and $\cos(x)$ in Eq.~(\ref{eq_key_Gloss}).
As one expects \cite{RubinsteinBook}, one obtains to leading order that
$\Gstor \to \Geq$, Eq.~(\ref{eq_Geqdef}), and $\Gloss \to \omega c_0$ 
as indicated by bold solid lines in Fig.~\ref{fig_GstorGloss} and Fig.~\ref{fig_GstorGloss_zeta}.
We have fitted $\Gloss$ for $\zeta=5$ using the independently determined value $c_0 \approx 16$. 

\paragraph*{Intermediate peaks.}
In the intermediate frequency regime one observes a striking peak at $\omega \approx 25$ 
for $\Gstor$, i.e. at the planewave frequency corresponding to the typical particle distance, 
Eq.~(\ref{eq_omegap}), and at a slightly smaller value $\omega \approx 15$ for $\Gloss$.
The phase angle $\delta$ characterizing the ratio $\Gloss/\Gstor \equiv \tan(\delta)$
is presented in the inset (in grad). It reveals a maximum at $\omega \approx 7$.

\paragraph*{High-frequency limit.}
In the high-frequency limit it becomes increasing difficult
to relax the imposed affine strain by non-affine displacements and, hence, to dissipate
the work done on the system. This implies that
\begin{equation}
\Gstor = \Geq + \Cstor \to (\muA-\muF) + \muF = \muA 
\label{eq_Gstor_olarge}
\end{equation}
for $\omega \to \infty$ (bold dashed horizontal lines), 
while $\Gloss$ decays with a non-universal (see below) sharp cutoff (dashed and dash-dotted lines).
This asymptotic behavior can be also understood on mathematical grounds by expanding 
Eq.~(\ref{eq_key_Gstor}) and Eq.~(\ref{eq_key_Gloss}) by integration by parts. 
This leads to two series-expansions in terms of even and odd derivatives of $C(t)$ taken at $t=0$ 
for, respectively, $\Gstor$ and $\Gloss$. We remind that $C(t)$ is an even function 
for classical systems \cite{AllenTildesleyBook,HansenBook} and that in a MD simulation
$C(t)$ must become analytic at sufficiently short times ($\dtMD < t \ll 1/\zeta$)
where thermostat effects are negligible. 
(Strict non-analyticity of $C(t)$ at $t=0$ is formally found, e.g., for the 
Maxwell model or the polymer Rouse model \cite{DoiEdwardsBook}.)
Analyticity at $t=0$ implies that the (converging) expansion of $\Gstor$ is dominated 
by its first term for $\omega \to \infty$. This demonstrates Eq.~(\ref{eq_Gstor_olarge}). 
The effective odd derivatives of $C(t)$ must, 
however, become negligible with decreasing friction $\zeta$, increasing angular frequency 
$\omega$ and decreasing time step $\dtMD$. Hence, $\Gloss$ must rapidly vanish. 
This point is further investigated in the final paragraph of this section.

\paragraph*{Friction effects.}
The friction effects presented in Fig.~\ref{fig_GstorGloss_zeta} are rather small for $\Gstor$
due to the large static constants $\Geq$ and $\muA$ dominating the storage modulus. 
As one would expect, they are more marked for the loss modulus $\Gloss$, especially at low frequencies where 
the thermostat has more time to dissipate the applied mechanical energy, but also at the large-$\omega$ cutoff. 
For $\zeta=5$ one might be tempted to fit an $1/\omega^3$-decay (dashed lines)
which indicates an {\em effective} non-analytical behavior of $C(t)$ at short times. 
With decreasing $\zeta$ the loss modulus is, however, better described by the exponential decay 
\begin{equation}
\Gloss = \muF \sqrt{\pi/4} \ x \ \exp(-x^2/2) \mbox{ with } x = \omega \thetaG
\label{eq_Gaussfit}
\end{equation}
as indicated by the dash-dotted line in panel (b) of Fig.~\ref{fig_GstorGloss_zeta}.
This phenomenological fit is the Cosine-Fourier transform assuming a Gaussian 
$C(t) = \muF \exp(-(t/\thetaG)^2/2)$ with the parameter $\thetaG \approx 0.1$ 
determined from the short-time (high-frequency) limit.
A similar phenomenological fit for the large-$\omega$ behavior of 
is also obtained by assuming a Lorentzian $C(t) = \muF/(1+(t/\thetaG)^2)$.
We note finally that if smaller time steps and larger frequencies could be computed, 
one expects even for higher friction constants a similar non-algebraic cutoff, 
albeit with a different parameter $\thetaG(\zeta)$.

\section{Conclusion}
\label{sec_conc}

\paragraph*{Summary.}
We have studied in the present work simple isotropic solids formed by permanent spring networks.
Some relevant static properties have been reviewed and characterized in Secs.~\ref{theo_static}, 
\ref{sec_refer} and \ref{simu_static}. More importantly, we have reconsidered theoretically 
(Sec.~\ref{sec_theo}) and numerically (Sec.~\ref{sec_simu}) the linear-response relation 
between the shear stress relaxation modulus $G(t)$ and 
the shear stress autocorrelation function $C(t)$ both in the time 
(Sec.~\ref{simu_key}, Figs.~\ref{fig_keyA}-\ref{fig_keyB}) and 
the frequency domain 
(Sec.~\ref{simu_oscil}, Figs.~\ref{fig_GstorGloss} and \ref{fig_GstorGloss_zeta}).
According to our key relations, Eqs.~(\ref{eq_key}-\ref{eq_key_Gloss}), 
$G(t)$ and $C(t)$ must become different in the solid limit  ($\Geq > 0$)
and it is thus impossible to determine $\Geq$ from $C(t)$ or its Fourier transforms 
$\Cstor$ and $\Closs$.
The short-time behavior of $C(t)$ and the large-$\omega$ limit of $\Cstor$ only yield the 
stress-fluctuation contribution $\muF$ to the equilibrium modulus $\Geq = \muA - \muF$ 
(Sec.~\ref{theo_static}) as already stressed elsewhere \cite{Mezard10,WXB15}.
The peak frequencies observed for $\Gstor$ and $\Gloss$ (Figs.~\ref{fig_GstorGloss} 
and \ref{fig_GstorGloss_zeta}) are qualitatively expected \cite{WTBL02,TWLB02} from 
the breakdown of continuum mechanics for large wavevectors $q \gg 1$ seen from the Fourier 
transformed transverse displacement field (Fig.~\ref{fig_static}).

\paragraph*{Discussion.}
It is obviously common and may often be helpful to describe a plateau of $C(t)$ 
at short or intermediate times 
(or equivalently at large or intermediate frequencies for $\Cstor$)
in terms of a finite shear modulus $\GM$ of a 
dynamical relaxation model, such as the Maxwell model 
for viscoelastic fluids or the reptation model of entangled polymer melts \cite{RubinsteinBook,DoiEdwardsBook}.
However, such a model allowing the theoretical {\em interpretation} of the data should not 
be confused with the proper {\em measurement procedure}, Eq.~(\ref{eq_Geqdef}),  
and a finite model parameter $\GM$ not with the equilibrium modulus $\Geq$ of the system 
which, incidentally, vanishes both for a Maxwell fluid or a linear polymer melt.  
In this sense different operational ``static" and ``dynamical" definitions of the shear modulus 
are used in the literature for describing glass-forming liquids close to the glass transition 
\cite{Klix12,Szamel11,Yoshino12,Mezard10,ZT13}. This may explain why qualitatively different 
temperature dependences ---
cusp singularity \cite{Mezard10,ZT13} {\em vs.} finite jump \cite{GoetzeBook,Klix12,Szamel11} ---
have been predicted recently. 
Hence, while our recent attempts to determine $\Geq(T)$ for two glass-forming model 
systems \cite{WXP13} are consistent with a continuous cusp, this is not necessarily 
in contradiction with a jump singularity for $\GM(T)$ determined from an intermediate
shoulder of $C(t)$ \cite{Klix12,Szamel15}.

\paragraph*{Outlook.}
We note finally that generalizing our key relations one obtains readily that
\begin{equation}
M(t) = C(t) + \Meq \ \mbox{ for }  t >0
\label{eq_final}
\end{equation}
for the relaxation modulus $M(t)$ of any continuous intensive variable $I$
with $\Meq = \partial I / \partial X$ being the equilibrium modulus
and $C(t) = \beta V \langle \delta \Ihat(t) \delta \Ihat(0) \rangle$
the corresponding autocorrelation function computed at a constraint
thermodynamically {\em conjugated} extensive variable $X$.
In the frequency domain this leads
to $\Mstor = \Meq + \Cstor$ and $\Mloss = \Closs$.
A natural example is provided by the relaxation modulus $M(t)=K(t)$ associated
to the normal pressure $I = P$. While $C(t)$ and $\Cstor$ must obviously vanish in the static limit for,
respectively, large times and small frequencies, $K(t)$ and $K^{\prime}(\omega)$ approach 
a finite compression modulus $\Meq=\Keq$ for stable (non-critical) thermodynamic systems.

\vspace*{0.2cm} 
\begin{acknowledgments}
H.X. thanks the IRTG Soft Matter for financial support.
We are indebted to H. Meyer (ICS, Strasbourg) and M. Fuchs (Konstanz) for helpful discussions.
\end{acknowledgments}


\begin{thebibliography}{44}
\bibitem[{foo({\natexlab{a}})}]{foot_gamnot}
\bibinfo{note}{For simplicity of the notation we write $\muF$ for $\muFgam$ and
  $C(t)$ for $\Ctgam$. We only note $\muFgam$ and $\muFtau$ or $\Ctgam$ and
  $\Cttau$ if fluctuations at imposed $\gamma$ (\NVgT-ensemble) and $\tau$
  (\NVtT-ensemble) are compared.}
\bibitem[{foo({\natexlab{c}})}]{foot_prime}
\bibinfo{note}{With the exception of the storage modulus $\Gstor$ and the loss
  modulus $\Gloss$ a prime denotes generally a derivative of a function with
  respect to its argument.}
\bibitem[{foo({\natexlab{e}})}]{foot_Berthier}
\bibinfo{note}{We emphasize that the dynamical Lebowitz-Percus-Verlet transform
  \cite{Lebowitz67} relies on the condition that only the distribution of start
  points of the trajectories depends on the ensemble, but not the relaxation
  pathways themselves \cite{AllenTildesleyBook}. While this does not hold for
  extensive variables, this is generally the case for fluctuations of
  instantaneous intensive variables. Interestingly, a similar approach based on
  Ref.~\cite{Lebowitz67} has been used for the four-point dynamic
  susceptibility $\chi_4(t)$ comparing its decay at constant temperature and
  constant energy \cite{Berthier05,Berthier07}.}
\end{thebibliography}

\end{document}